\documentclass[twocolumn]{aastex631}

\usepackage{amsmath}
\usepackage{footmisc}
\usepackage{rotating,longtable}

\usepackage{graphicx}
\usepackage{subfigure}
\usepackage{natbib} 
\usepackage{hyperref}
\usepackage{tablefootnote}
\usepackage{booktabs}

\usepackage{lipsum,mwe}
\newcommand{\kms}{\,km\,s$^{-1}$}

\shorttitle{[HKS2019] E70}
\shortauthors{Verma et al.}

\graphicspath{{./}{fig/}}

\begin{document}

\title{Exploring Stellar Cluster and Feedback-driven Star Formation in Galactic Mid-infrared Bubble [HKS2019] E70}

\correspondingauthor{Aayushi Verma}
\email{aayushiverma@aries.res.in}

\author[0000-0002-6586-936X]{Aayushi Verma}
\affiliation{Aryabhatta Research Institute of observational sciencES (ARIES), Manora Peak, Nainital-263001, India}

\author[0000-0001-5731-3057]{Saurabh Sharma}
\affiliation{Aryabhatta Research Institute of observational sciencES (ARIES), Manora Peak, Nainital-263001, India}
\author[0000-0002-3873-6449]{Kshitiz K. Mallick}
\affiliation{Aryabhatta Research Institute of observational sciencES (ARIES), Manora Peak, Nainital-263001, India}
\author[0000-0001-6725-0483]{Lokesh Dewangan}
\affil{Physical Research Laboratory, Navrangpura, Ahmedabad—380 009, India} 
\author[0000-0001-9312-3816]{Devendra K. Ojha}
 \affil{ Department of Astronomy and Astrophysics, Tata Institute of Fundamental Research (TIFR), Mumbai 400005, Maharashtra, India}

\author[0000-0002-6740-7425]{Ram Kesh Yadav}
\affiliation{National Astronomical Research Institute of Thailand (NARIT), Chiang Mai 50200, Thailand}

\author[0000-0002-7485-8283]{Rakesh Pandey}
\affiliation{Physical Research Laboratory, Navrangpura, Ahmedabad—379 009, India}

\author[0000-0001-7650-1870]{Arpan Ghosh}
\affiliation{Aryabhatta Research Institute of observational sciencES (ARIES), Manora Peak, Nainital-263001, India}

\author{Harmeen Kaur}
\affiliation{Center of Advanced Study, Department of Physics DSB Campus, Kumaun University Nainital, 263001}

\author[0000-0002-0151-2361]{Neelam Panwar}
\affiliation{Aryabhatta Research Institute of observational sciencES (ARIES), Manora Peak, Nainital-263001, India}

\author{Tarak Chand}
\affiliation{Aryabhatta Research Institute of observational sciencES (ARIES), Manora Peak, Nainital-263001, India}

\begin{abstract}
We present a comprehensive analysis of the Galactic mid-infrared (MIR) bubble [HKS2019] E70 (E70) by adopting a multi-wavelength approach to understand the physical environment and star formation scenario around it. We identified a small (radius $\sim$1.7 pc) stellar clustering inside the E70 bubble and its distance is estimated as 3.26 $\pm$ 0.45 kpc. This cluster is embedded in the molecular cloud and hosts massive stars as well as young stellar objects (YSOs), suggesting active star formation in the region. The spectral type of the brightest star `M1' of the E70 cluster is estimated as O9V and a circular ring/shell of gas and dust is found around it. The diffuse radio emission inside this ring/shell, the excess pressure exerted by the massive star `M1' at the YSOs core, and the distribution of photo-dissociation regions (PDRs), a Class $\textsc{i}$ YSO, and two ultra-compact (UC) H\,{\sc ii} regions on the rim of this ring/shell, clearly suggest positive feedback of the massive star `M1' in the region. 
We also found a low-density shell-like structure in $^{12}$CO(J=1-0) molecular emission along the perimeter of the E70 bubble. The velocity structure of the $^{12}$CO emission 
suggests that the feedback from the massive star appears to have expelled the molecular material 
and subsequent swept-up material is what appears as the E70 bubble.

\end{abstract}


\section{Introduction} \label{sec:intro}

Active star-forming regions in molecular clouds generally consist of young star clusters (YSCs), bubbles, clouds/filaments, and massive star/s. Bubbles are one of the most fascinating objects noticed within modern large-scale infrared (IR) surveys (e.g., \citealt{2006ApJ...649..759C,2010AJ....139.2330W,2019PASJ...71....6H}). These are shell-like structures created by the interaction of expanding H\,{\sc ii} region with its neighboring interstellar medium, which provide a target to explore the effects of massive stellar feedback on its surrounding material and led to exploring how an expanding H\,{\sc ii} region can trigger the formation of a new generation of stars (e.g. \citealt{2010A&A...523A...6D,2012ApJ...755...71K,2019MNRAS.487.1517L,2020ApJ...897...74Z}).

\begin{figure*}
    \centering
    \includegraphics[width=0.45\textwidth]{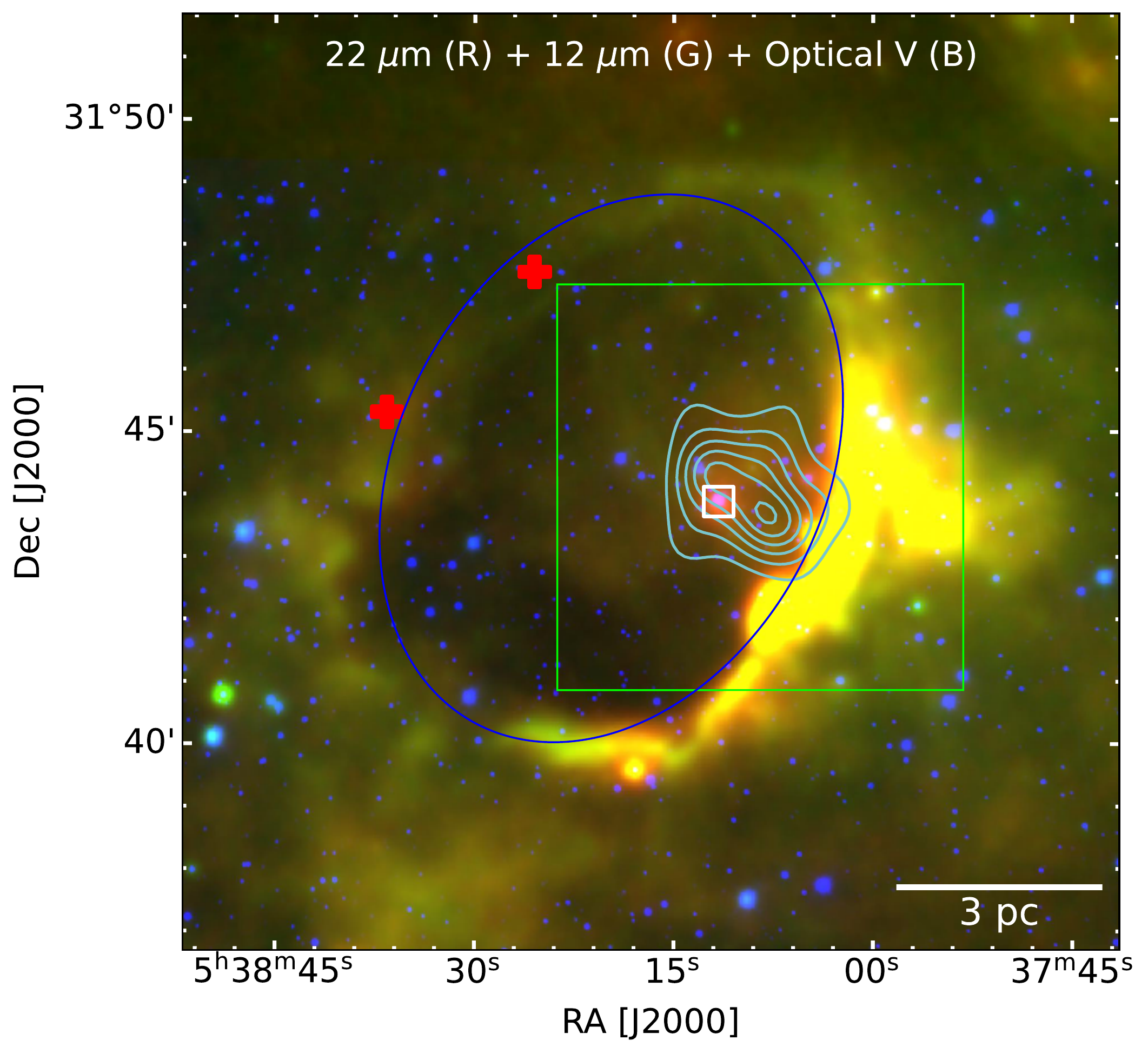}
    \includegraphics[width=0.45\textwidth]{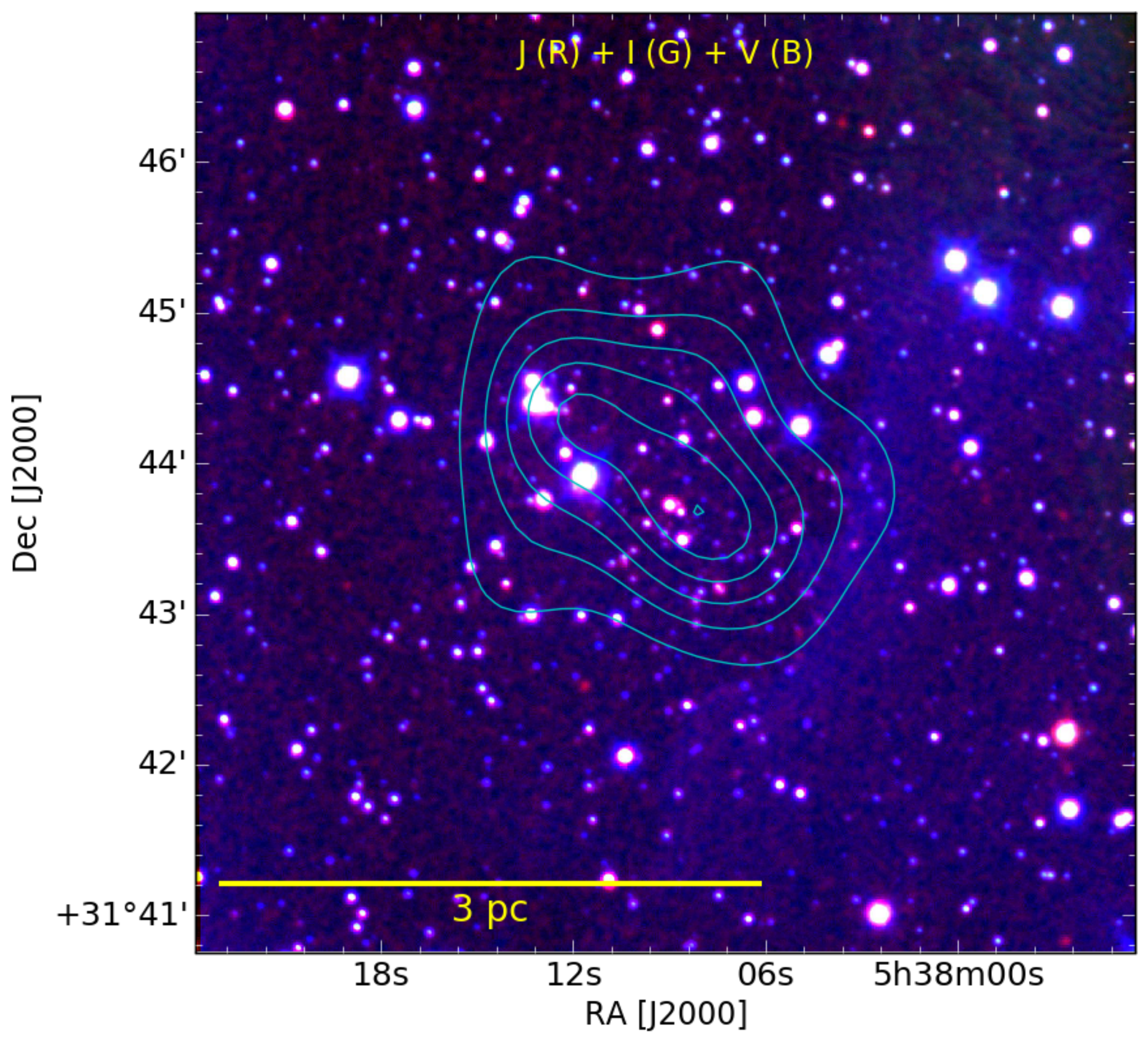}
    \caption{Left panel: Color-composite image (Red: \emph{WISE} 22 $\mu$m; Green: \emph{WISE} 12 $\mu$m; and Blue: DFOT $V$ band) of $15^{\prime} \times 15^{\prime}$ region overlaid with isodensity contours (cyan) generated from the NIR catalog (cf. Section \ref{sec:clustering}). The lowest level for the isodensity contours is 30 stars arcmin$^{-2}$ with a step size of 8 stars arcmin$^{-2}$. The green square represents the $6^\prime.5 \times6^\prime.5$ region which has been selected to observe using 3.6m DOT whereas the blue ellipse represents the elliptical-shape of the E70 bubble. The locations of probable UC H\,{\sc ii} regions and `M1' are also marked with red `$+$' and white square, respectively. Right Panel: Color-composite image (Red: 2MASS $J$ 1.25 $\mu$m; Green: DOT $I_c$ band; and Blue: DOT $V$ band) of the zoomed-in view of green square region (of left panel) overlaid with the isodensity contours (cyan) for E70 derived using the NIR catalog.} \label{fig:intro_rgb}
\end{figure*}

Most of the IR bubbles own ionizing massive stars ($\geq8\,M_\odot$) somewhere near the centers of polycyclic aromatic hydrocarbons (PAH) shells creating an H\,{\sc ii} region which eventually triggers the formation of stars. However, the exact mechanism behind triggered star formation and its contribution to the overall star formation rate remains uncertain. 
For the present study, we have chosen a rarely studied Galactic mid-infrared (MIR) bubble [HKS2019] E70 (hereafter E70; $\alpha_{J2000}=05^h38^m19^{s}$.1 and $\delta_{J2000}=+31^{\circ}44\arcmin20\arcsec$) that was first cataloged by \citet{2019PASJ...71....6H} using the \emph{AKARI} and \emph{Herschel} photometric data. This bubble is reported to be located at a distance of 3.3 $\pm$ 0.4 kpc and has a radius of 4$\arcmin$.48 with a covering fraction of 0.92 \citep{2019PASJ...71....6H}. The left panel of Figure \ref{fig:intro_rgb} represents the color-composite image (Red: \emph{WISE} 22 $\mu$m; Green: \emph{WISE} 12 $\mu$m, and Blue: Optical $V$ band taken using 1.3m Devasthal Fast Optical Telescope (DFOT)) of the E70 bubble region. An elliptical-shaped ring of dust and gas at 12 $\mu$m is visible (represented by the blue ellipse) on the map. The eccentricity of this ellipse is estimated as $\sim0.7$.  The length of the semi-major and semi-minor axes are $\sim4^\prime.6$ and $\sim3^\prime.4$, respectively.
The \emph{WISE} 12 $\mu$m hints towards PAH emission featuring at 11.3 $\mu$m. These emissions are strong indications of the photo-dissociation regions (PDRs) generated due to the effect of feedback from the massive star \citep{2020ApJ...896...29K, 2020ApJ...905...61P}. We noticed that a bright star located inside the rim of the bubble (`M1', marked with a white square) coincides with the peak 22 $\mu$m emission. We anticipate that this might be a massive star responsible for creating the ring/shell of warmed-up gas and dust. 
We also noticed diffuse radio emission (an indicator of ionized gas) inside this ring/shell along with two very young probable ultra compact (UC) H\,{\sc ii} regions on the ring/shell, from the archival NVSS data (marked by red `+' in the left panel of Figure \ref{fig:intro_rgb}). Thus, the E70 bubble region is an interesting site to investigate the role of massive stars in creating the bubble of gas and dust and subsequently the formation of a new generation of stars.

In this paper, we did a detailed multi-wavelength analysis of the E70 bubble region. The structure of the paper is as follows: Section \ref{sec:observation} describes the observations and data reduction as well as archival data sets employed in our analysis. The stellar density profile in this region, membership probability, derivation of fundamental parameters (i.e., age and distance) of the region, mass function (MF) analyses, physical environment around the E70 bubble, surrounding molecular cloud morphology,  etc. are presented in the Section \ref{sec:result}. The results from our analysis are discussed in Section \ref{sec:discussion}, and we sum up our study in Section \ref{sec:conclusion}.

\section{Observation and data reduction}\label{sec:observation}

\subsection{Optical Photometric Observation and Data Reduction}

The deep optical photometric observations of the E70 bubble were carried out in $VI_c$ bands on 2022 February 1, using 4K$\times$4K CCD IMAGER having a field of view (FOV) of $6^\prime.5 \times6^\prime.5$ which is mounted at the axial port of 3.6m  Devasthal Optical Telescope (DOT), Nainital \citep{2018BSRSL..87...29K,2022JApA...43...27K}. 
The green square in the left panel of Figure \ref{fig:intro_rgb} represents $6^\prime.5 \times6^\prime.5$ region which has been selected to observe using 3.6m DOT and the right panel of Figure \ref{fig:intro_rgb} is a representation of the zoomed-in view of this region via a color-composite image (Red: 2MASS $J$ 1.25 $\mu$m; Green: DOT optical $I_c$ band; and Blue: DOT optical $V$ band) overlaid with isodensity contours (please refer Section 3.2).
The images were taken in 2 $\times$ 2 binning mode for a total integration time of 115 minutes and 70 minutes in $V$ and $I_c$ bands, respectively. The readout noise and gain for the CCD are 10 e$^-$ s$^{-1}$ and 5 e$^-$ ADU$^{-1}$, respectively. 
Calibration of the instrumental magnitudes from the 3.6m DOT observations was done by secondary standard stars generated from the observations of the same region and a standard field \citep[SA95,][]{1992AJ....104..340L} by using the 1.3m DFOT, Nainital, on 2021 October 13, in broadband $U$, $B$, $V$, and $I_c$ filters using 2K $\times$ 2K CCD camera having a FOV of $18\arcmin.5 \times18\arcmin.5$ \citep{2012ASInC...4..173S}.  
Several flat and bias frames were also taken during the observations.

The basic data reduction including image cleaning, photometry, and astrometry, was done using the standard procedure explained in \citet{2020MNRAS.498.2309S}. We transformed instrumental magnitudes into standard Vega systems by using the following transformation equations  \citep[cf.][]{1992ASPC...25..297S}.

\begin{equation}
    \begin{split}
        u& = U + (4.691 \pm 0.006)\\
        &- (0.093 \pm 0.005)(U-B) + (0.611 \pm 0.006)X_U,
    \end{split}
\end{equation}

\begin{equation}
    \begin{split}
        b& = B + (2.900 \pm 0.005)\\
        &- (0.126 \pm 0.003)(B-V) + (0.284 \pm 0.009)X_B,
    \end{split}
\end{equation}

\begin{equation}
    \begin{split}
        v& = V + (2.290 \pm 0.003)\\
        &+ (0.101 \pm 0.002)(V-I_c) + (0.130 \pm 0.005)X_V,
    \end{split}
\end{equation}
\\and\\
\begin{equation}
    \begin{split}
        i_c& = I_c + (2.516 \pm 0.007)\\
        &- (0.061 \pm 0.004)(V-I_c) + (0.058 \pm 0.014)X_I.
    \end{split}
\end{equation}

\begin{figure}
    \centering
    \includegraphics[width=0.48\textwidth]{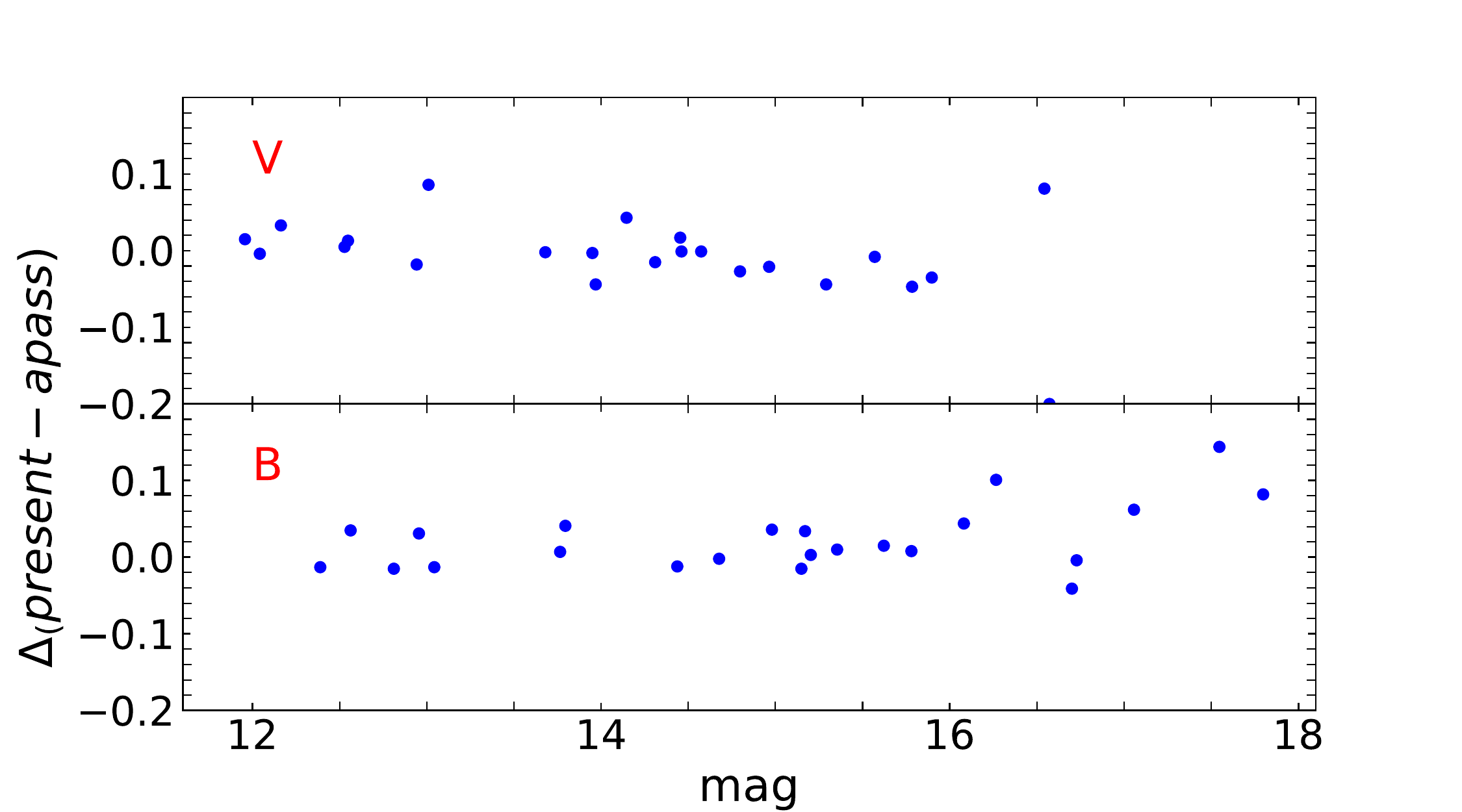}
    \includegraphics[width=0.48\textwidth]{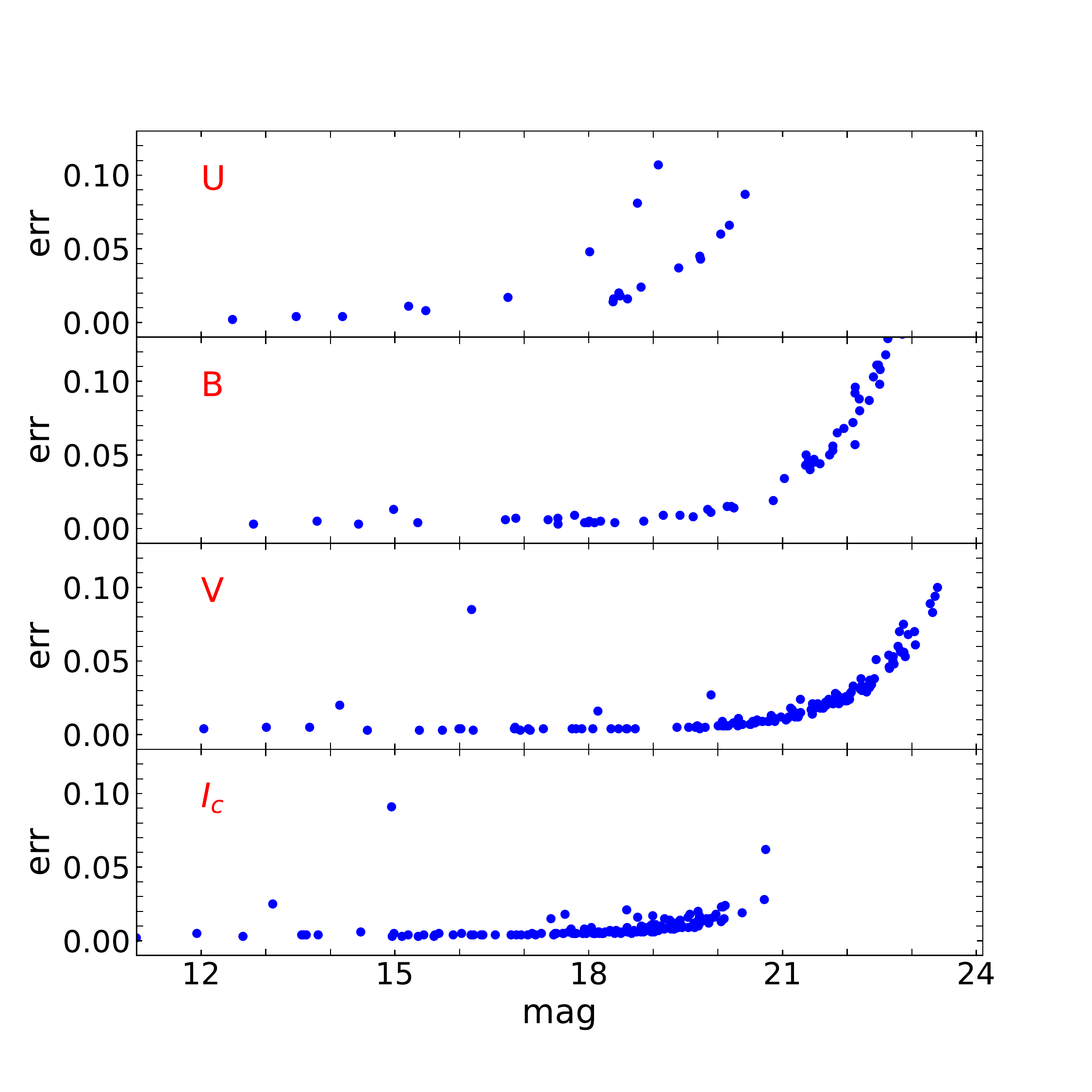}
    \caption{ Upper panel: Comparison of the present photometry with “APASS” in the $V$ and $B$ bands for E70. Lower panel: Photometric error vs. magnitude plots in $U$, $B$, $V$, and $I_c$ filters. }
    \label{fig:optical_mag}
\end{figure}

Here, $U$, $B$, $V$, and $I_c$ represent the standard magnitudes, and $u$, $b$, $v$, and $i_c$ represent the instrumental magnitudes which are normalized for the corresponding exposure time; and X's represent the air mass in respective bands. We compared the present standard magnitudes with archive `APASS'\footnote{The AAVSO Photometric All-Sky Survey, https://www.aavso.org/apass} (cf. Figure \ref{fig:optical_mag} upper panel) and found no shifts in our calibrated magnitudes. 
The stars were identified with detection limits (photometric error $\leq0.1$, cf. Figure  \ref{fig:optical_mag} lower panel) of 21.76, 24.34, 23.64, 21.20 mags in $U$, $B$, $V$, and $I_c$ bands, respectively.
We used Graphical Astronomy and Image Analysis Tool\footnote{http://star-www.dur.ac.uk/~pdraper/gaia/gaia.html} to do the astrometry of the stars with rms noise of the order of $\sim0\arcsec.3$.

\subsection{Completeness of Photometric Data}\label{sec:completeness}
\begin{figure}
    \centering
    \includegraphics[width=0.48\textwidth]{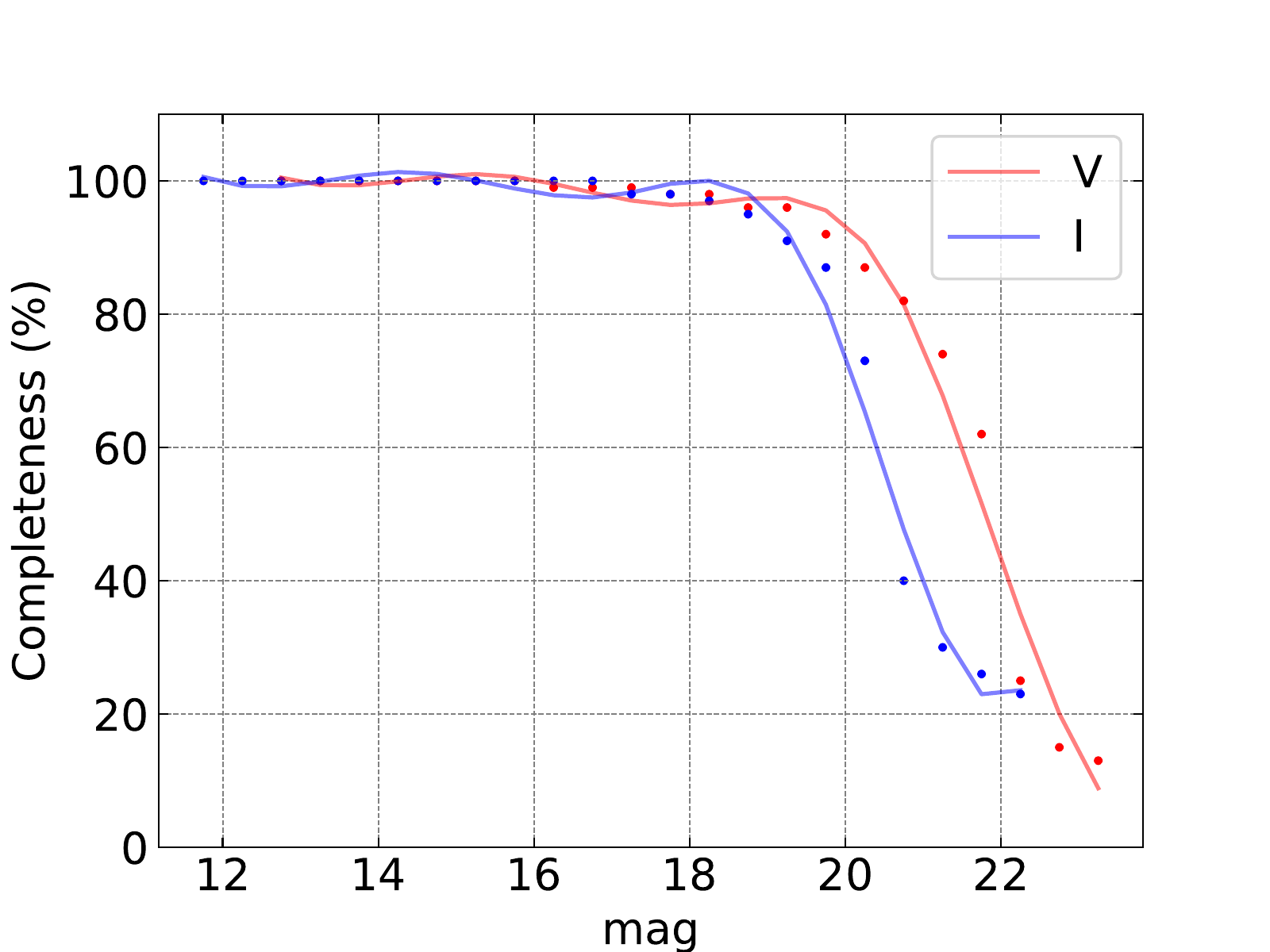}
    \caption{Completeness factor as a function of magnitude derived using the artificial star experiments (ADDSTAR).}
    \label{fig:completeness}
\end{figure}

Various factors cause the incompleteness of the photometric data, such as nebulosity, detection limit, stars' crowding, etc. In that case, it becomes crucial to evaluate the completeness limit of the data. To find out the completeness factor (CF), we used the {\sc addstar} routine of IRAF (as described in \citealt{2008AJ....135.1934S}). In a nutshell, a few artificial stars of known magnitudes and positions are added to the original frames randomly and the generated frames are then again reduced by the same procedure as that of the original frames. CF is obtained as a function of magnitude by the ratio of the number of stars recovered to those added in each magnitude interval and is shown in Figure \ref{fig:completeness}. As expected, the incompleteness of the photometric data increases as we go towards fainter limits and the stars up to $V\sim21.0$  and $I_c\sim20.0$  are detected with CF $\ge 75\%$. 

\begin{deluxetable*}{lll}
\tablecolumns{3}
\tablewidth{0pt}
\tabletypesize{\small}
\caption{Details of uGMRT Observations}
\tablehead{
\colhead{} & \colhead{610 MHz} & \colhead{1400 MHz}}
\startdata
Date of Observation & 2022 Sep 30   & 2022 Oct 02 \\
Phase Center & $\alpha_{2000}$=05$^h$38$^m$19$^s$.1 & $\alpha_{2000}$=05$^h$38$^m$19$^s$.1 \\
             & $\delta_{2000}$=31$^{\circ}$44\arcmin22\arcsec.1 & $\delta_{2000}$=31$^{\circ}$44\arcmin20\arcsec.0 \\
Flux Calibrator   &  3C147    & 3C147 \\
Phase Calibrator  &  0535+391 & 0535+391 \\
Cont. Bandwidth   &  32 MHz   & 32 MHz \\
Primary Beam      &  23\arcmin & 38\arcmin \\
\hline
Synthesised Beam\tablenotemark{\small{a}}   &  5\arcsec.08 $\times$ 3\arcsec.65 &  5\arcsec $\times$ 4\arcsec\, (2\arcsec.09 $\times$ 1\arcsec.51) \\
rms noise ($\mu$Jy/beam)  &  $\sim44$ & $\sim15$ ($\sim9$) \\
\hline
Northern UC H\,{\sc ii}\tablenotemark{\small{a},\small{b},\small{c}}    &   & \\
\cmidrule(lr){1-1}
~~ Size          &  5\arcsec.2 $\times$ 3\arcsec.8  &  5\arcsec.8 $\times$ 4\arcsec.4\, (2\arcsec.7 $\times$ 1\arcsec.7)               \\
~~ Integral Intensity (mJy)  &  3.72 $\pm$ 0.08      & 1.94 $\pm$ 0.03 (1.68 $\pm$ 0.02) \\
~~ Peak Intensity (mJy\,beam$^{-1}$) &  3.48 $\pm$ 0.04      & 1.53 $\pm$ 0.02 (1.13 $\pm$ 0.01) \\
\hline
Southern UC H\,{\sc ii}\tablenotemark{\small{a},\small{c}}    &   & \\
\cmidrule(lr){1-1}
~~ Size          &  5\arcsec.3 $\times$ 3\arcsec.9  &  6\arcsec.0 $\times$ 4\arcsec.6\, (2\arcsec.9 $\times$ 1\arcsec.9)               \\
~~ Integral Intensity (mJy)  &  4.71 $\pm$ 0.08      & 3.34 $\pm$ 0.03 (2.75 $\pm$ 0.02) \\
~~ Peak Intensity (mJy\,beam$^{-1}$) &  4.21 $\pm$ 0.04      & 2.46 $\pm$ 0.02 (1.58 $\pm$ 0.01) \\
\enddata
\label{table_GMRTObs}
\tablenotetext{a}{The values in brackets for 1400\,MHz are for the highest
resolution map which could be made.}
\tablenotetext{b}{Only the main peak (See Figure \ref{fig_GMRTMaps}) was fit.}
\tablenotetext{c}{From fitting a Gaussian using the AIPS task {\sc jmfit}.}
\end{deluxetable*}

\subsection{Optical Spectroscopic Observation and Data Reduction}

The optical spectroscopic observations of a probable massive star `M1', showing surrounding warm dust emission (cf. Figures \ref{fig:intro_rgb}), were carried out on 2022 December 12, using the Hanle Faint Object Spectrograph Camera (HFOSC) instrument mounted on the 2 m Himalayan Chandra Telescope (HCT), Hanle, India. The observations were performed with GRISM 7 (3800–6840 \r{A}) with a resolution power (R) of $\sim$ 1200. A FeAr calibration lamp was also observed during the same night.

The spectroscopic data were reduced with IRAF packages following the procedure outlined in \citet{2012MNRAS.424.2486J}. Aperture extraction, line identification using lamp and dispersion correction were achieved by {\sc apall, identify}, and {\sc dispcor} tasks, respectively. Eventually, normalization of the spectrum was undertaken using the {\sc continuum} task.

\subsection{Radio Continuum Observation}

Low-frequency radio continuum observations were carried out for this region using uGMRT (upgraded Giant Metrewave Radio Telescope) radio
interferometric array \citep{SwarupGMRT_91, Gupta_ugmrt_2017CSci} at 610\,MHz and 1400\,MHz (PI: Aayushi Verma, Proposal ID: 42\_089). Table \ref{table_GMRTObs} shows the details of the observations. For the purpose of our analysis, we utilized the GSB (GMRT Software Backend) data which corresponds to the legacy GMRT bandwidth of 32\,MHz\footnote{https://www.gmrt.ncra.tifr.res.in/doc/GMRT\_specs.pdf}.

Data reduction for the 610\,MHz band was carried out using the SPAM (Source Peeling and Atmospheric Modelling) pipeline \citep{Intema_SPAM_2009AA,Intema_SPAM_2014ASInC,Intema_TGSS_2017AA}. The automated pipeline is a Python module that interfaces with the Astronomical Image Processing System (AIPS) software. Since at lower frequencies, the primary beam size is large (and thus a larger FOV), the SPAM pipeline carries out direction-dependent (ionospheric) calibration and imaging. The final image from the pipeline includes primary beam correction and system temperature correction using the 408\,MHz map of \citet{Haslam408MHz_1982} \citep[see][for details]{Intema_TGSS_2017AA}. Since the SPAM pipeline is only for sub-GHz bands, data reduction for 1400\,MHz was carried out separately using the tasks in AIPS software, broadly involved the following main steps \citep[also see][]{Mallick_I16148_2015MNRAS} : multiple iterations of flagging and calibration, averaging and splitting the source uvdata after applying the calibrations, and facet imaging with a few rounds of self-calibration. The uvdata so obtained after the final round of self-calibration was again used for (facet) imaging, and the resulting facets were flattened into a final image. Primary beam correction was applied to this image via the AIPS task {\sc pbcor}\footnote{coefficients from http://www.ncra.tifr.res.in/ncra/gmrt/gmrt-users/observing-help/}. As this source is on the Galactic plane, the image was then re-scaled by a factor calculated using the (408\,MHz) sky temperature map of \citet{Haslam408MHz_1982} \citep[see][for more details]{Mallick_I16148_2015MNRAS}. The resultant radio maps were then used for analysis.

\subsection{Molecular line data}
\label{section_MolecularCOData}

We obtained the archival 30\arcmin $\times$ 30\arcmin\, $^{12}$CO$(J=1-0)$, $^{13}$CO$(J=1-0)$, and C$^{18}$O$(J=1-0)$ molecular line observations which have been observed as a part of MWISP (Milky Way Imaging Scroll Painting) project by the PMO (Purple Mountain Observatory) 13.7\,m millimeter-wave radio telescope (\citealt{ 633694461037117445}; \citealt{Su_MWISP_2019ApJS}).
While the $^{12}$CO spectral cube has a channel width of $\sim$ 0.16 \kms, the same for the $^{13}$CO and C$^{18}$O cubes is about 0.17 \kms. The spatial resolution and grid size for all the cubes are $\sim$\,50\arcsec\, and 30\arcsec, respectively. The rms noise was calculated to be $\sim$\,0.46\,K ($^{12}$CO), $\sim$\,0.23\,K ($^{13}$CO), and $\sim$\,0.23\,K (C$^{18}$O), where the pixel brightness is in T$_{\mathrm{MB}}$ (main beam temperature) units.

\subsection{Other Archival Data}

\begin{table*}[]
    \footnotesize
    \centering
    \caption{List of miscellaneous surveys endorsed for the present study (NIR to Radio Wavelength)}
    \begin{tabular}{c c c c}
    \hline
    Survey & Wavelength/s & $\sim$ Resolution & Reference\\
    \hline
    Two Micron All Sky Survey\footnote{\citet{https://doi.org/10.26131/irsa2}} (2MASS) & 1.25, 1.65, and 2.17 $\mu$m & $2\arcsec.5$ & \citet{2006AJ....131.1163S}\\
    \emph{Gaia} DR3\footnote{https://www.cosmos.esa.int/web/gaia/dr3} (magnitudes, parallax, and PM) & 330–1050 nm & 0.4 mas & \citet{2022arXiv220800211G}\\
    \emph{Herschel} Infrared Galactic Plane Survey\footnote{http://archives.esac.esa.int/hsa/whsa/} & 70, 160, 250, 350, 500 $\mu$m & $5\arcsec.8$, $12\arcsec$, $18\arcsec$, $25\arcsec$, $37\arcsec$ & \citet{2010PASP..122..314M}\\
    NRAO VLA Sky Survey\footnote{https://www.cv.nrao.edu/nvss/postage.shtml} (NVSS) & 21 cm & $46\arcsec$ & \citet{1998AJ....115.1693C}\\
    \emph{Spitzer} GLIMPSE360 Survey\footnote{\citet{https://doi.org/10.26131/irsa214}} & 3.6, 4.5 $\mu$m & $2\arcsec$, $2\arcsec$ & \citet{2005ApJ...630L.149B}\\
    Wide-field Infrared Survey Explorer\footnote{\citet{https://doi.org/10.26131/irsa1}} (\emph{WISE}) & 3.4, 4.6, 12, 22 $\mu$m & $6\arcsec.1$, $6\arcsec.4$, $6\arcsec.5$, $1\arcsec$ & \citet{2010AJ....140.1868W}\\
    UKIRT InfraRed Deep Sky Survey\footnote{http://wsa.roe.ac.uk/} (UKIDSS) & 1.25, 1.65, and 2.22 $\mu$m & $0\arcsec.8$, $0\arcsec.8$, $0\arcsec.8$ & \citet{2008MNRAS.391..136L}\\
    Milky Way Imaging Scroll Painting (MWISP)  & $^{13}$CO(J =1$-$0),  $^{12}$CO(J =1$-$0)  &  $50\arcsec$ & \citet[]{Su_MWISP_2019ApJS} \\
    \hline
    \end{tabular}
    \label{tab:archival_data}
\end{table*}

We used various archival data sets ranging from NIR to radio for our target source. A brief specification of these data sets is given in Table \ref{tab:archival_data}. The \emph{Herschel} column density and temperature maps (spatial resolution $\sim12\arcsec$) have been downloaded directly from the publicly available website\footnote{http://astro.cardiff.ac.uk/research/ViaLactea/}. These maps are procured for EU-funded ViaLactea project \citep{2010PASP..122..314M} adopting the Bayesian PPMAP technique \citep{2010A&A...518L.100M} at 70, 160, 250, 350, and 500 $\mu$m wavelengths \emph{Herschel} data \citep{2015MNRAS.454.4282M, 2017MNRAS.471.2730M}.

\section{Result and Analysis}\label{sec:result}

\subsection{Spectral Analysis of `M1'}

\begin{figure*}
    \centering
    \includegraphics[width=0.98\textwidth]{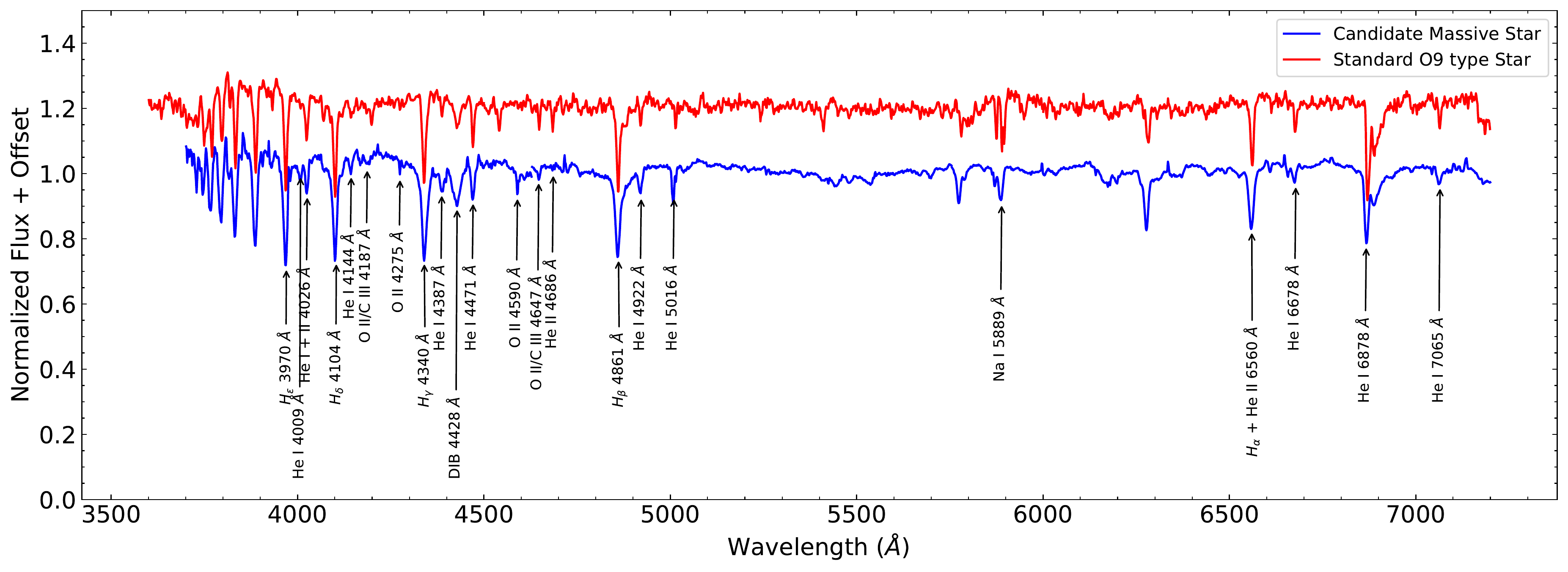}
    \caption{Wavelength-calibrated normalized spectrum of `M1' shown with blue color. We have plotted the standard spectrum of an O9V-type star (red color) from \citet{1984ApJS...56..257J} for comparison purposes. The observed spectral lines are also marked.}
    \label{fig:optical_spectra}
\end{figure*}

The obtained wavelength-calibrated normalized spectrum of `M1' is presented in Figure \ref{fig:optical_spectra} with blue color. The signal-to-noise ratio\footnote{This has been calculated using the \emph{specutils} package of Python} of this spectrum is $\sim$ 277 \citep{nicholas_earl_2023_7803739}. We utilized various criteria and spectral libraries available in the literature (cf. \citealt{1984ApJS...56..257J,1990PASP..102..379W}) for the spectral classification. The Spectra of O and B-type stars have hydrogen, helium, and other atomic lines (e.g. C\,{\sc iii}, Mg\,{\sc ii}, O\,{\sc ii}, Si\,{\sc iii}, Si\,{\sc iv}). The ratio of He\,{\sc i} 4471 \r{A}/He\,{\sc ii} 4542 \r{A} is a key indicator of the spectral type. If the spectral type of the star is later than O7, then it is greater than 1. For late O-type stars, the line strength of He\,{\sc ii} gradually gets weaker and is last noticed in B0.5-type stars \citep{1990PASP..102..379W}. If the absorption line of He\,{\sc ii} is observed at 4200 \r{A} along with O\,{\sc ii}/C\,{\sc iii} line at 4650 \r{A}, then the spectral type of the star is earlier than B1.

In the spectrum, we noticed prominent hydrogen lines (at 3970, 4104, 4340, 4861, and 6560 \r{A}) along with the neutral helium lines (at 4009, 4026, 4144, 4387, 4471, 4922, 5016, 6678, 6878, and 7065 \r{A}) and ionized helium lines (at 4026, 4686 and 6560 \r{A}). In addition to them, we also observed O\,{\sc ii} line at 4590 \r{A}, O\,{\sc ii}/C\,{\sc iii} line at 4187 and 4647 \r{A} and Na line at 5889 \r{A}.

We did the spectral classification of `M1' by a visual comparison of its spectrum with the standard library spectra described in \citet{1984ApJS...56..257J}. We find that this candidate has a spectral type O9V with an uncertainty of $\pm$1 in the classification of subclass and hence it is a massive star ($\geq8 M_\odot$).

\subsection{Structure and Stellar Clustering in the E70 Bubble}\label{sec:clustering}

To understand the structure and stellar clustering in the E70 bubble, we generated the stellar surface density maps by nearest neighbor (NN) method (as given in \citealt{2005ApJ...632..397G}) on NIR catalog (cf. Appendix \ref{app:yso_identification}). We varied the radial distance in such a way that encompasses the twentieth nearest star with a grid size of 6$\arcsec$. The local stellar surface density $\sigma$ at each grid position [i, j], is computed by:
\begin{equation}
    \sigma(i,j) = \frac{N}{\pi r_N^2(i,j)}
\end{equation}

Here $r_N^2(i,j)$ represents the projected radial distance to the \emph{N$^{th}$} nearest star.

We overlaid the estimated isodensity contours (cyan curves) in Figure \ref{fig:intro_rgb}. The lowest level for isodensity contours is 30 stars arcmin$^{-2}$ with a step size of 8 stars arcmin$^{-2}$. A clear clustering of stars having a bit of elliptical geometry can be seen inside the ring of dust and gas at $\alpha_{J2000}=05^h38^m07^s$.7 and $\delta_{J2000}=+31^{\circ}43\arcmin38\arcsec$.
Due to this elliptical morphology, we redefine the area of this cluster through its \emph{convex hull}\footnote{\emph{Convex hull} is a polygon enclosing all points in a grouping with internal angles between two contiguous sides of less than 180$^{\circ}$.} instead of its circular area which generally overestimates the area of an elongated cluster (\citealt{2006A&A...449..151S}, \citealt{2016AJ....151..126S}, \citealt{2020MNRAS.498.2309S}). 
The area of the cluster ($A_{cluster}$ = 16.86 arcmin$^2$) is estimated from the area of the convex hull ($A_{hull}$) normalized by a geometric factor \citep[cf. for details,][]{2020MNRAS.498.2309S}, given as:

\begin{equation}
    \begin{split}
        A_{cluster}=\frac{A_{hull}}{1-\frac{n_{hull}}{n_{total}}}
    \end{split}
\end{equation}

where, $n_{hull}$ is the total number of vertices on the hull and $n_{total}$ is the total number of points inside the hull. The \emph{convex hull} is generated from the position of stars located inside the lowest isodensity contour (cf. upper right panel of Figure \ref{fig:yso_extinction}).

The aspect ratio $\frac{R^2_{circ}}{R^2_{cluster}}$
of this cluster comes out to be 0.91. The radius of the cluster $R_{cluster}$ (= 1$\arcmin$.80 = 1.70 pc) is the radius of the circle whose area is equal to $A_{cluster}$. $R_{circ}$ (= 1$\arcmin$.74 = 1.65 pc) is defined as half of the farthest distance between two hull objects.

\subsection{Extinction, Distance, and Age of the E70 bubble}

\begin{figure*}
    \centering
    \includegraphics[width=0.45\textwidth]{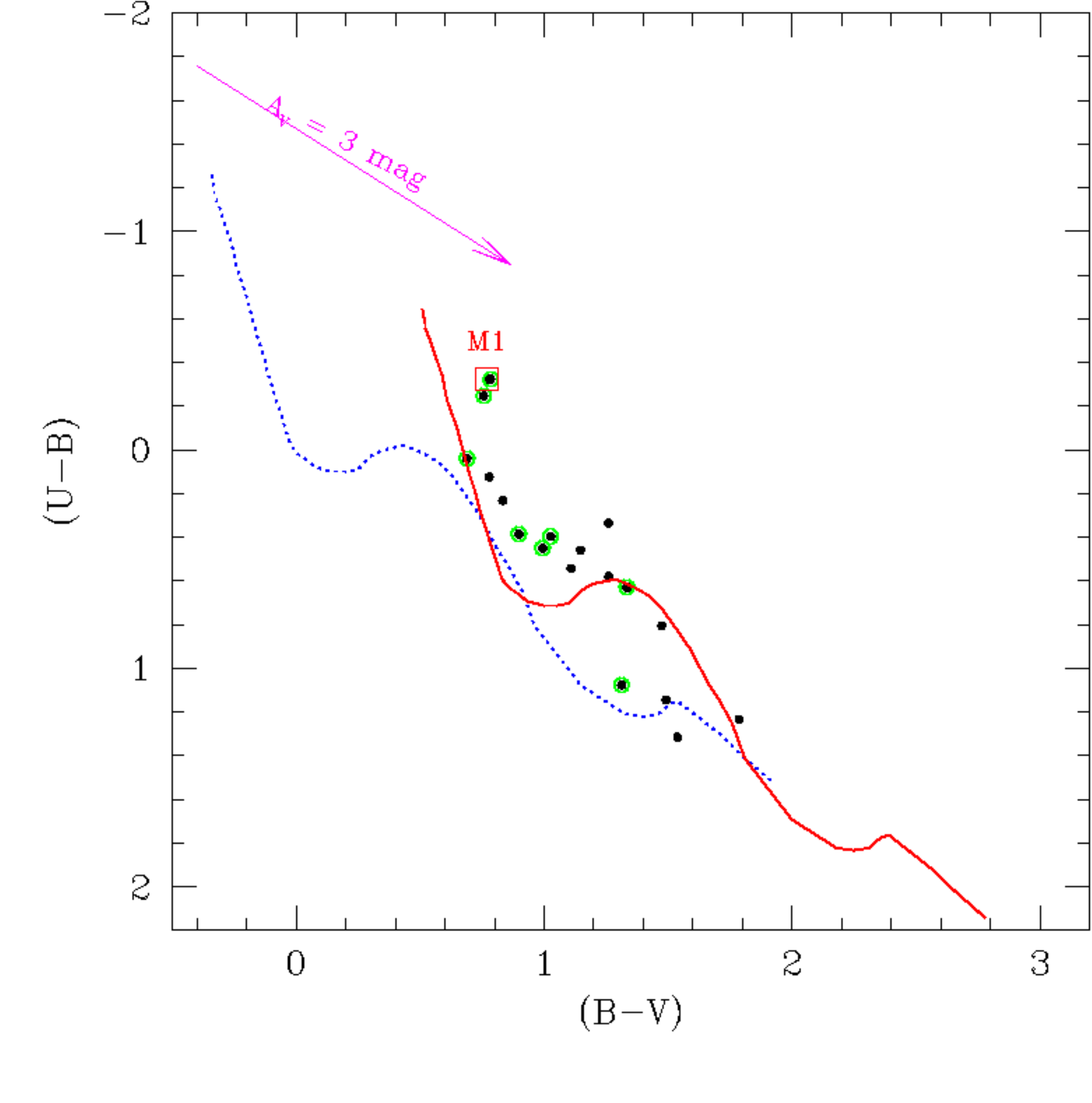}
    \includegraphics[width=0.45\textwidth]{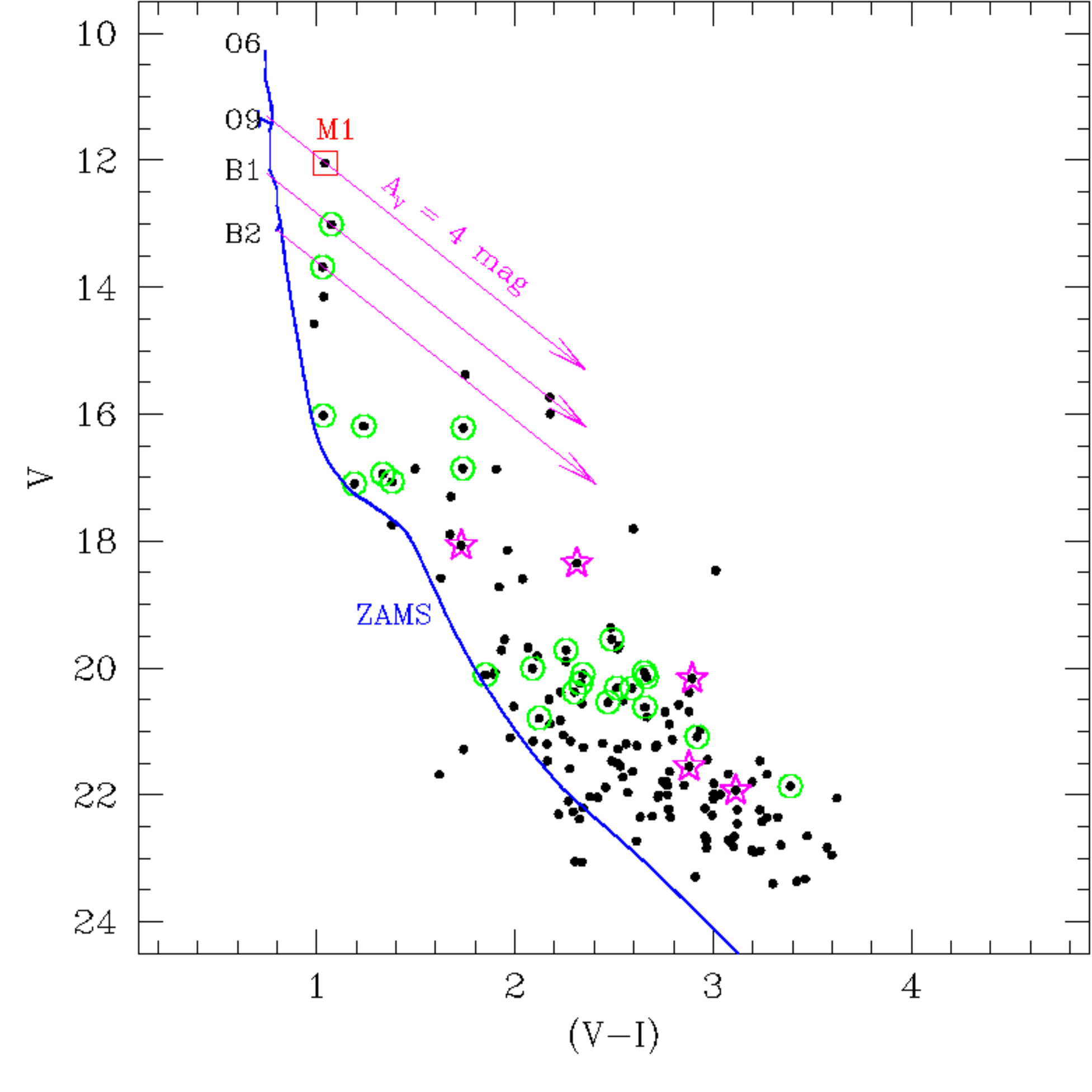}
    \caption{Left panel: $(U-B)$ vs. $(B-V)$ TCD for the sources in the E70 region. The \emph{Gaia DR3} members are also plotted with green rings. The dotted blue curve is representing the intrinsic ZAMS for $Z=0.02$ given by 
    \citet{2013ApJS..208....9P}.
    The continuous red curve is representing the ZAMS which is shifted along the reddening vector. Right panel: $V$ vs. $(V-I_c)$ CMD. The blue solid curve is representing the ZAMS isochrone by \citet{2013ApJS..208....9P}, corrected for the distance 3.26 kpc and reddening $E(B-V) = 0.85$ mag. The PMS stars are shown by asterisks. The massive star `M1' is marked on both panels by a red square symbol.}
    \label{fig:optical_tcd}
\end{figure*}

We used $(U-B)$ vs. $(B-V)$ two-color diagram (TCD) for the estimation of extinction towards the E70 bubble, as shown in the left panel of Figure \ref{fig:optical_tcd}. We show the distribution of stars located in the E70 cluster, i.e., those lying within the convex hull shown in Figure \ref{fig:yso_extinction},
using the black dots along with the intrinsic zero-age main sequence (ZAMS; dotted blue curve) which we took from \citet{2013ApJS..208....9P}. 
We have also over-plotted the distribution of probable member stars (identified on the basis of PM data, cf. Appendix \ref{app:membership_prob}) by green circles. As this region contain the distribution of gas and dust, we expect differential reddening in this bubble which is also evident from the broad distribution of stars in the TCD. Therefore, the minimum reddening towards the bubble can be estimated by visually fitting the ZAMS to the bluer end of the distribution stars of spectral type A or earlier. Several factors, such as the distribution of binary stars, pre-main-sequence (PMS) stars, metallicity, and photometric errors; have also led us to such choices (see \citealt{1994ApJS...90...31P} for further details). We shifted the ZAMS along the reddening vector having a slope of $E(U-B)/E(B-V)$ = 0.72 (for normal Galactic reddening law `$R_V$ = 3.1', \citealt{1979A&AS...38..197M,1989AJ.....98..611G}) to the distribution of stars (red curve in the left panel of Figure \ref{fig:optical_tcd}), thus estimated minimum reddening towards the cluster as $E(B-V)_{min}$ = 0.85 $\pm$ 0.05 mag (corresponding to $A_V=2.64\pm0.20$ mag, for normal Galactic reddening law `$R_V$ = 3.1', \citealt{1979A&AS...38..197M,1989AJ.....98..611G}). The approximate error in the reddening measurement is determined by the procedure outlined in \citet{1994ApJS...90...31P}. The massive star `M1' seems to suffer more ($E(B-V)$ $\sim$ 0.5 mag) extinction above the minimum extinction value derived for the E70 cluster. 

To estimate the distance of this bubble, we used the mean of the distances reported by \citet{2021AJ....161..147B} of the 10 cluster members (see Appendix \ref{app:membership_prob} for detailed analysis on cluster membership), having membership probability $\geq 80 \%$ and parallax values with error $\leq 0.1$ mas. Thus, the distance of this cluster comes out to be 3.26 $\pm$ 0.45 kpc. Furthermore, we utilized  V vs. $(V-I_c)$ color-magnitude diagram (CMD) from our deep optical observations (see Figure \ref{fig:optical_tcd}, right panel) of the stars within the E70 cluster to confirm this estimated distance as well as to derive the age of E70 cluster itself (\citealt{2020MNRAS.492.2446P, 2020ApJ...891...81P, 2020ApJ...896...29K}).
The massive star `M1', probable members of the E70 cluster, and the identified PMS stars (cf. Appendix \ref{app:yso_identification}) are also marked by red squares, green circles, and magenta asterisks, respectively, in the right panel of Figure \ref{fig:optical_tcd}.
The intrinsic ZAMS (blue curve) which we took from \citet{2013ApJS..208....9P} is also plotted for an extinction $E(B-V)_{min}$ = $0.85$ mag and distance of 3.26 kpc. The ZAMS seems to be very well fitted to the blue envelope of the distribution of stars, thus confirming our distance and extinction estimates (for more detail on CMD isochrones fitting, refer \citealt{1994ApJS...90...31P}).  The location of the massive star `M1' in the CMD confirms its spectral type (O9V) as estimated by its optical spectrum. As this is the brightest star in the cluster region, the upper age limit of the E70 cluster should be $\sim$8.1 Myr \citep{2004fost.book.....S}.
The E70 cluster also seems to host more massive (B1-B2) stars. 
As some of the member stars fall in the PMS stage in the CMD and some of the stars are showing excess IR emission (stars with discs around them, age $<$ 3 Myr \citep{Evans_2009}), we can safely conclude that this region is still forming young stars even-though the most massive star was born $\sim$ 8.1 Myr ago.

\subsection{Mass function}\label{sec:mf}

\begin{figure}
    \centering
    \includegraphics[width=0.47\textwidth, angle=0]{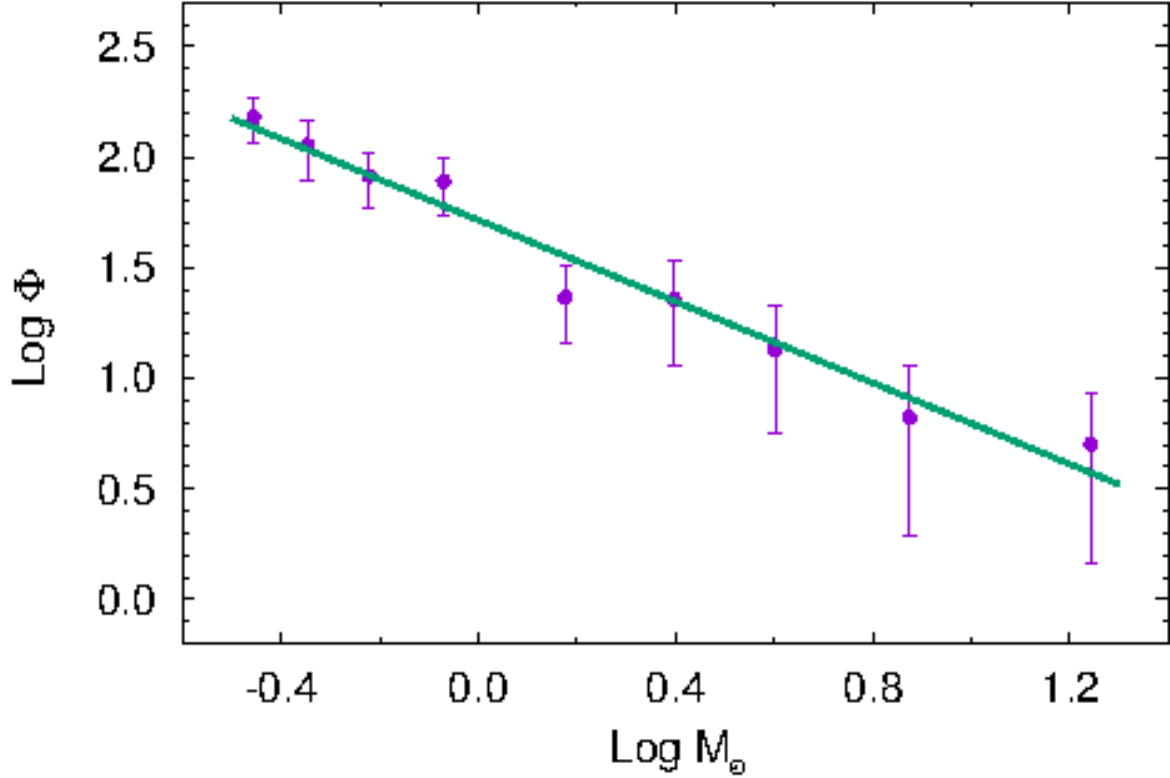}
    \caption {A plot of the MF for the cluster region using deep optical data taken from 3.6m DOT. Log $\phi$ represents log($N$/dlog $m$). The error bars represent $\pm\sqrt N$ errors. The solid line shows the least squares fit to the MF distribution (filled circles).}
    \label{mf}
\end{figure}

The initial mass function (IMF), defined as the distribution of stellar masses which are formed in a single star formation event, is an important statistical tool to understand star formation in a particular volume of space. The MF is usually expressed by a power law, given as $N(log\,m) \propto m^{\Gamma}$. The slope of MF is expressed as
\begin{equation}
    \Gamma = \frac{d\,log\,[N\,(log\,m)]}{d\,log\,m}.
\end{equation}

Here, N\,(log\,m) expresses the total number of stars per unit logarithmic mass interval. The MS is obtained with the help of $V$ versus $(V-I_c)$ CMD generated from the deep optical photometric data taken from the 3.6m DOT (cf. Figure \ref{fig:optical_tcd}) which is corrected for the data incompleteness (cf. Section \ref{sec:completeness}). The removal of the field star contamination and the estimation of the mass of the individual stars in the optical CMD is done using the procedure outlined in  \citet{2020ApJ...891...81P} and \citet{2017MNRAS.467.2943S}. The resultant MF distribution for the cluster region is shown in Figure \ref{mf}. The slope of the MF ($\Gamma$) in the mass range $\sim$ 0.3 $<$ M/M$_\odot$ $<$ 20 comes out to be $-0.92 \pm 0.06$ for the stars in the  E70 cluster region.

The higher-mass stars mostly follow the Salpeter MF \citep{1955ApJ...121..161S}. At lower masses, the IMF is less well-constrained, but appears to flatten
below 1 M$_\odot$  and exhibits fewer stars of the lowest masses \citep{2002Sci...295...82K, 2003PASP..115..763C, 2015arXiv151101118L, 2016ApJ...827...52L}.
In this study, we do not find the change in MF slope in the mass range $\sim$ 0.3 $<$ M/M$_\odot$ $<$ 20.

\subsection{Extraction of the young population embedded in the molecular cloud}\label{sec:extraction_of_cores}

\begin{figure*}[!t]
    \centering
    \includegraphics[width=0.45\textwidth]{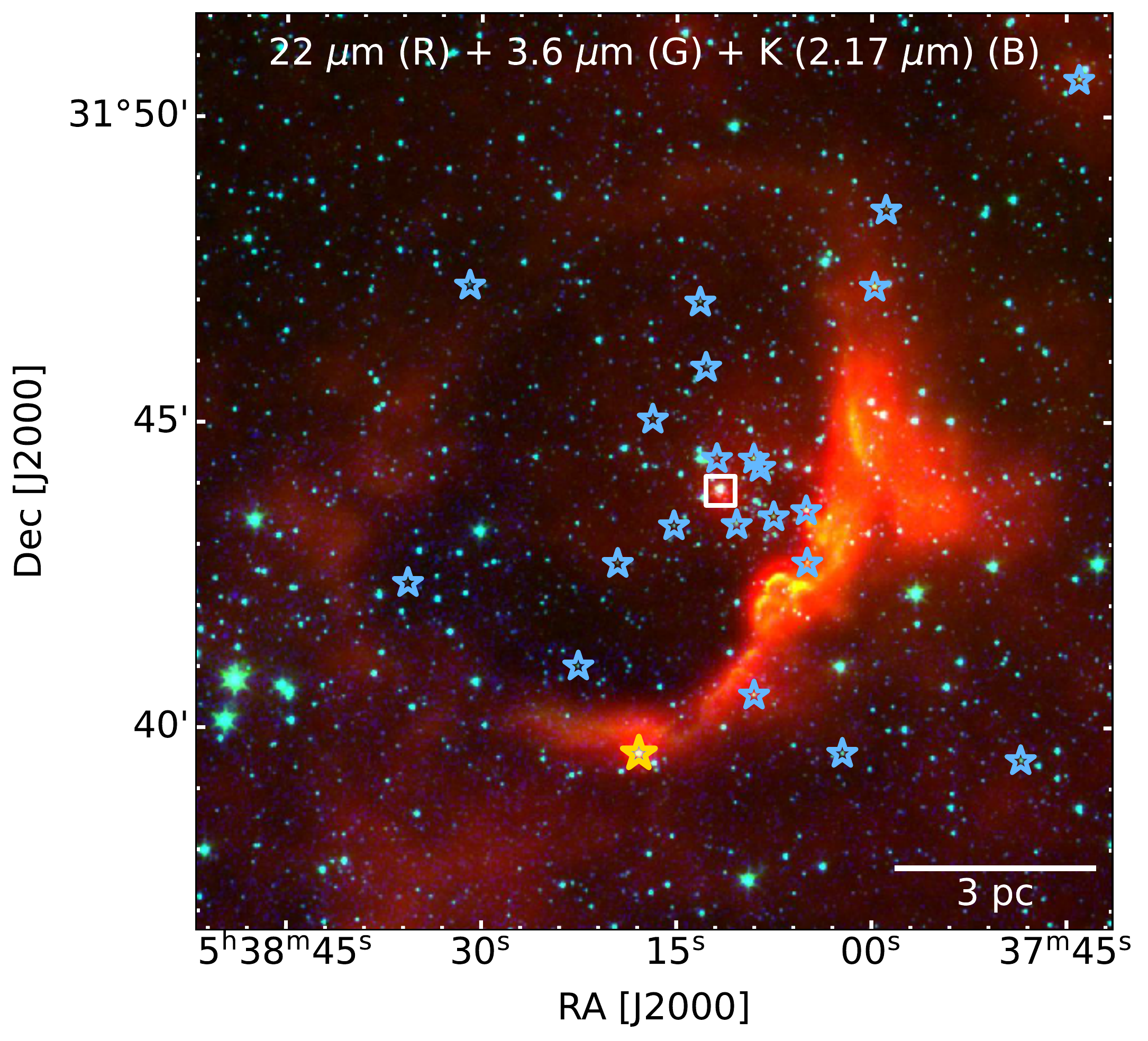}
    \includegraphics[width=0.45\textwidth]{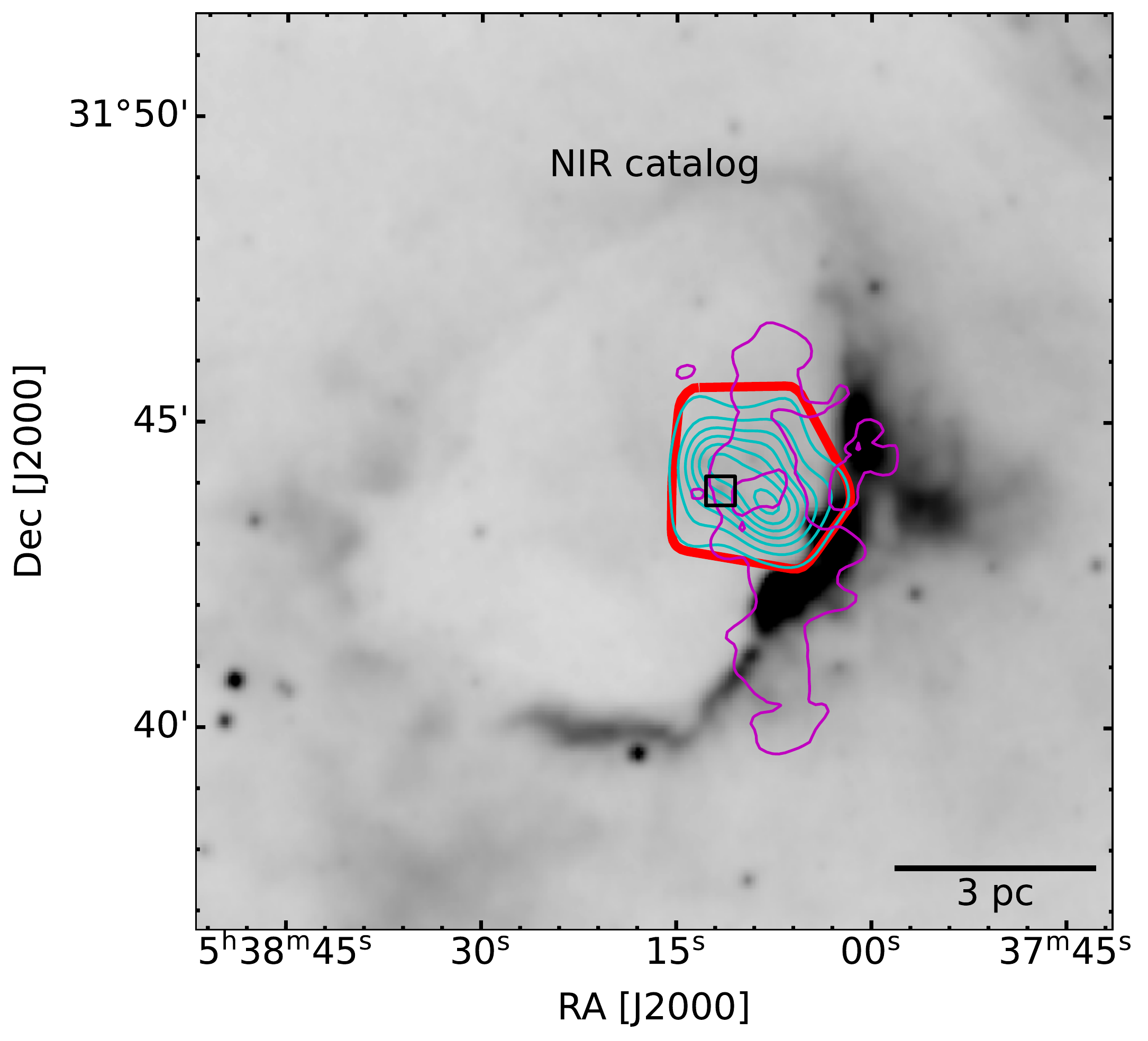}
    \includegraphics[width=0.45\textwidth]{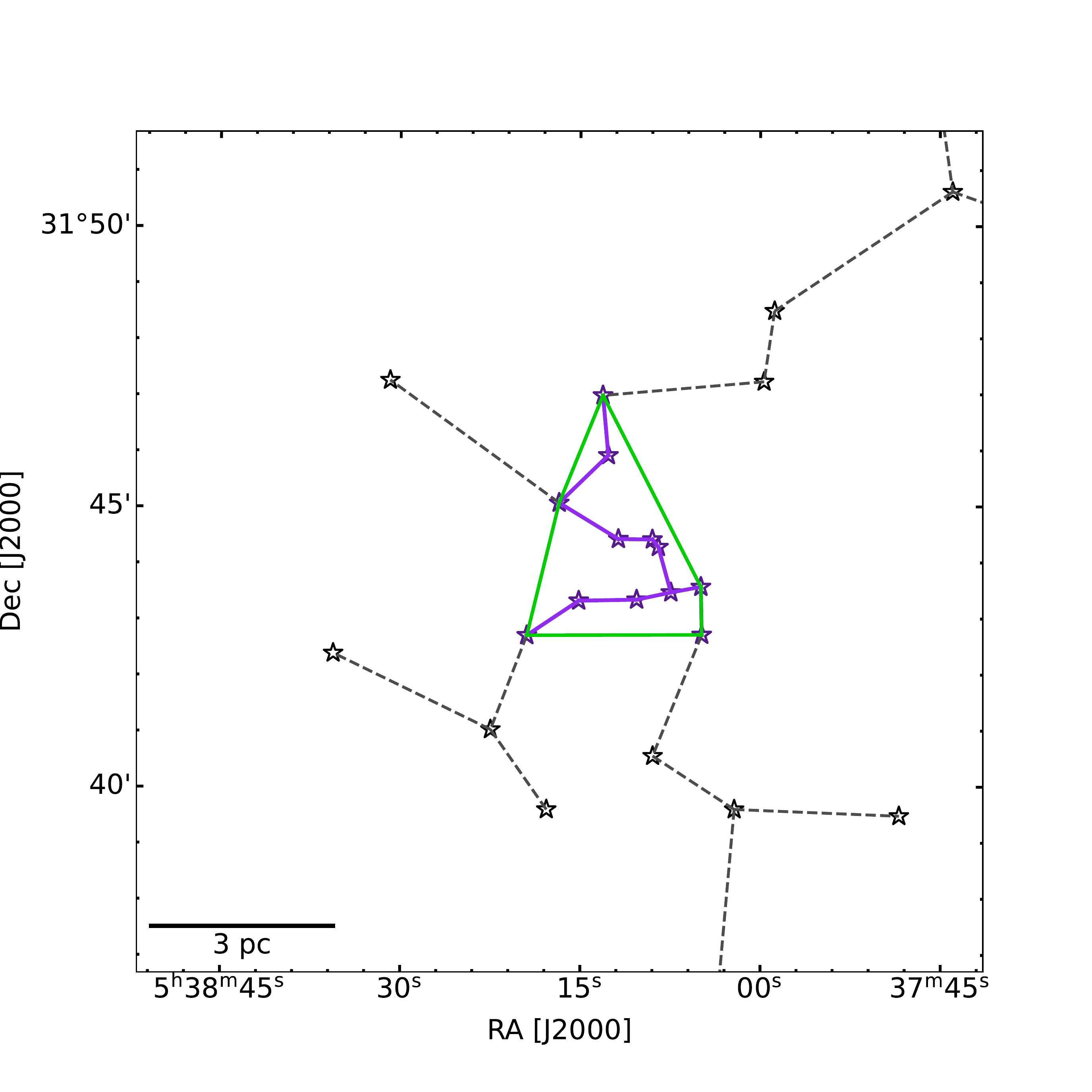}
    \includegraphics[width=0.43\textwidth]{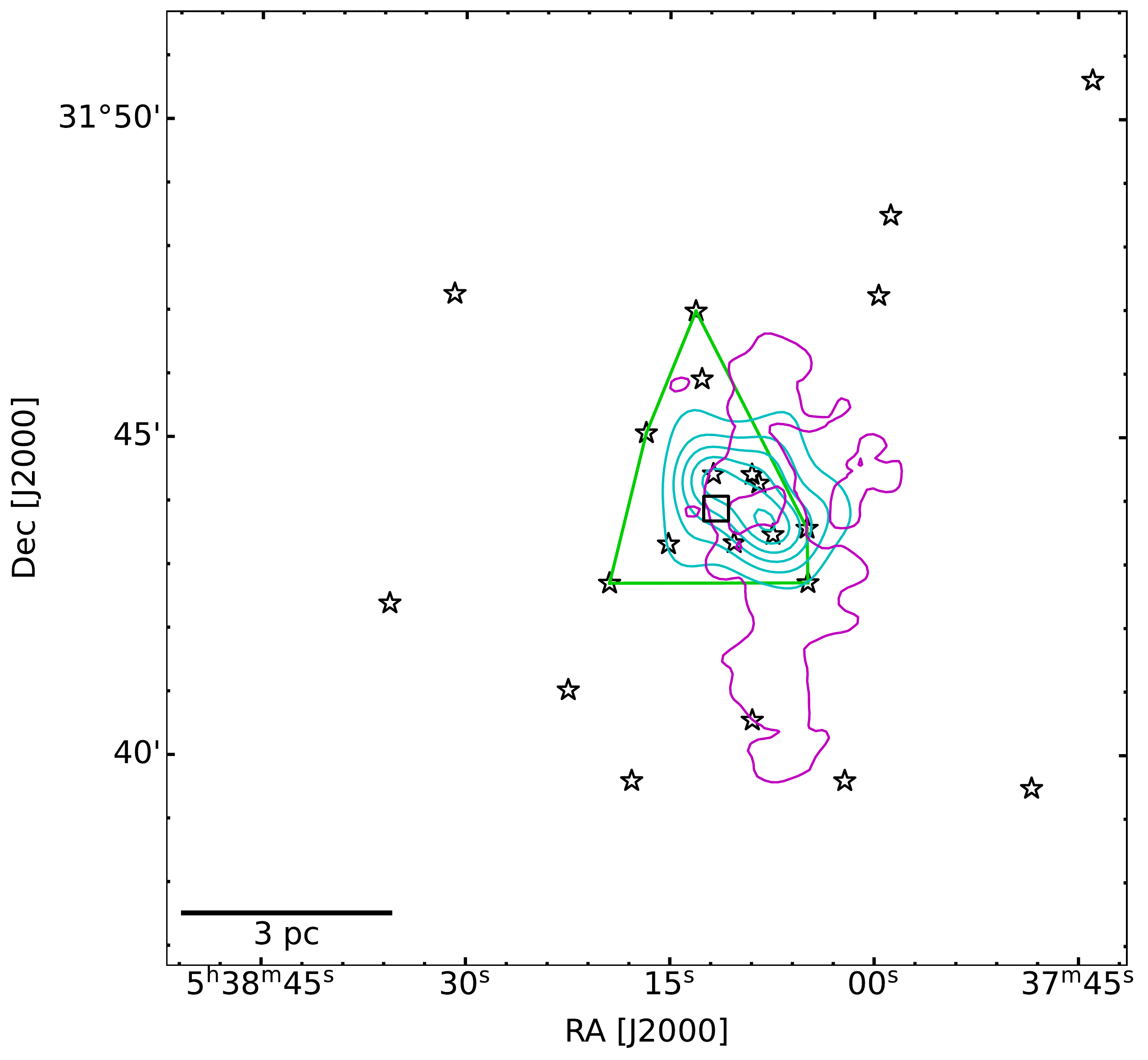}
    \caption{Upper left panel: Color-composite image (Red: \emph{WISE} 22 $\mu$m; Green: \emph{Spitzer} 3.6 $\mu$m and 2MASS K 2.17 $\mu$m) of $15^\prime \times15^\prime$ region overlaid with the locations of Class $\textsc{i}$ (yellow asterisk) and Class $\textsc{ii}$ (blue asterisk) YSOs. The location of the identified massive star `M1' is also marked with a white square. Upper right panel: \emph{Convex hull} (red) of the generated isodensity contours (cyan) along with the extinction map (magenta contours) generated after the removal of foreground stars, overlaid over \emph{WISE} 12 $\mu$m image of the E70 bubble. The lowest level for the extinction map is 2.7 mag with a step size of 1.9 mag. The black square symbol marks the location of the massive star `M1'.
    Lower left panel: MST connections in the core (purple) along with the location of YSOs (purple asterisk) in the core. The isolated core is also enclosed by the \emph{Convex hull} using green line segments respectively. The location of all the identified YSOs are also represented by black asterisks. Lower right panel: Extinction map (magenta contours) overlaid with the isodensity contours (cyan) and core region (green lines) along with the locations of identified YSOs (black asterisk) and massive star `M1' (black square).}
    \label{fig:yso_extinction}
\end{figure*}

The E70 bubble is showing many signatures of star-forming activities and thus, we can expect young populations of stars embedded in this region. Since these young populations (or young stellar objects (YSOs)) are usually associated with the circumstellar disc, we identified/classified them based on their excess IR-emission (see Appendix \ref{app:yso_identification} for details). We found 22 YSOs (1 Class $\textsc{i}$ and 21 Class $\textsc{ii}$) in our selected FOV of this region and their locations are shown in the top left panel of Figure \ref{fig:yso_extinction}.  
This panel shows the color-composite image generated using the \emph{WISE} 22 $\mu$m, \emph{Spitzer} 3.6 $\mu$m, and 2MASS K (2.17 $\mu$m) band images of the region overlaid with the location of identified massive star `M1' (white square).
The \emph{WISE} 22 $\mu$m gives an indication of the distribution of warm dust emission due to feedback from massive stars, whereas \emph{WISE} 3.6 $\mu$m image consists of PAH bands at 3.3 $\mu$m formed at PDRs under the influence of strong UV radiation from the massive stars.  We can easily see a bubble of gas and dust surrounding the massive stars where the western arc seems to be more brightened up (or have PDRs) in comparison to other parts of the bubble. There are also globules-like structures in the western arc of the bubble. 

In the upper-right panel of Figure \ref{fig:yso_extinction}, we show the isodensity contours generated from the NIR catalog and the convex hull for the stars located in the outermost isodensity contour overlaid over \emph{WISE} 22 $\mu$m image of the E70 bubble. We also show the distribution of extinction contours generated from the NIR catalog as explained in \citet{2009ApJS..184...18G}. The massive star `M1' (black square) is located within the convex hull of the cluster star. This cluster also seems to be associated with a high extinction region (or molecular cloud) as suggested by the extinction contours. The YSOs distribution is more extended than the cluster but is lying mostly within the MIR bubble.

To isolate the YSOs sharing similar star formation history, we applied an empirical technique \emph{`Minimal Spanning Tree'} (MST; \citealt{2009ApJS..184...18G}).
It is one of the best techniques as it isolates the groupings without any bias or smoothing and preserves underlying geometry \citep{2004MNRAS.348..589C,2006A&A...449..151S,2009MNRAS.392..868B,2009ApJS..184...18G}. The methodology for the extraction of MST is discussed in our previous publications, i.e., \citealt{2016AJ....151..126S, 2020ApJ...891...81P}, and is plotted in the bottom left panel of the Figure \ref{fig:yso_extinction}. We successfully isolated the grouping/core of YSOs (shown by the magenta asterisk) identified in this region from a diffuse distribution of YSOs (shown by black asterisks). We also generated the \emph{Convex hull} for the core members of the YSOs and shown by a green polygon in the bottom left panel of Figure \ref{fig:yso_extinction}.

In the bottom right panel of Figure \ref{fig:yso_extinction}, we show the distribution of the YSOs in the region along with the identified YSOs core which is enclosed by their \emph{Convex hull} (green polygon). We also over-plotted the isodensity and extinction contours in the figure. Clearly, the YSOs core and the E70 cluster are embedded in the high extinction region.

\subsection{Physical environment around E70 bubble}\label{sec:physical_environment}

\begin{figure*}
    \centering
    \includegraphics[width=0.98\textwidth]{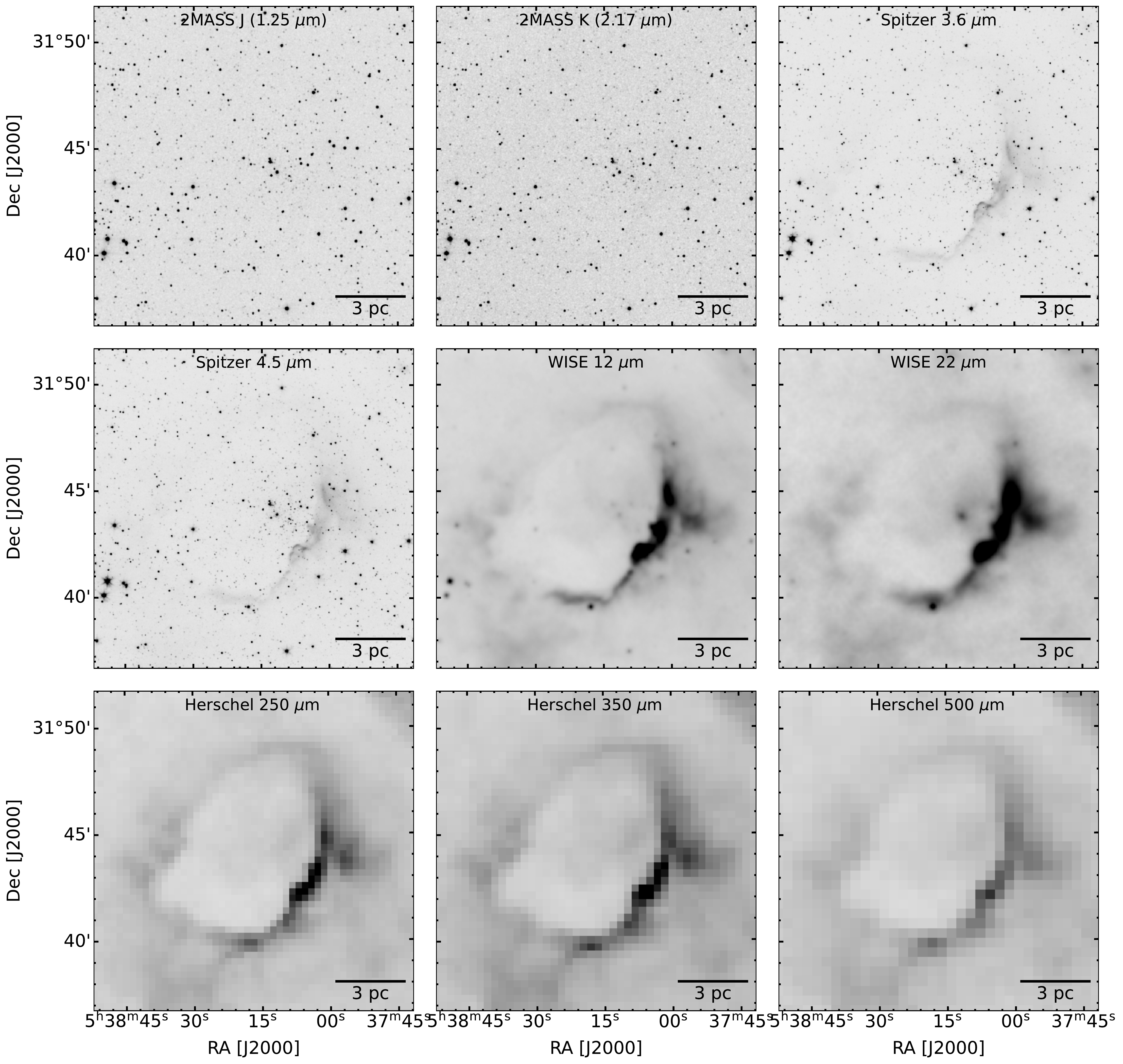}
    \caption{Multi-band images of the E70 bubble ranging from NIR to FIR wavelengths.}
    \label{fig:multiband}
\end{figure*} 

\begin{figure*}[!t]
    \centering
    \includegraphics[width=0.45\textwidth]{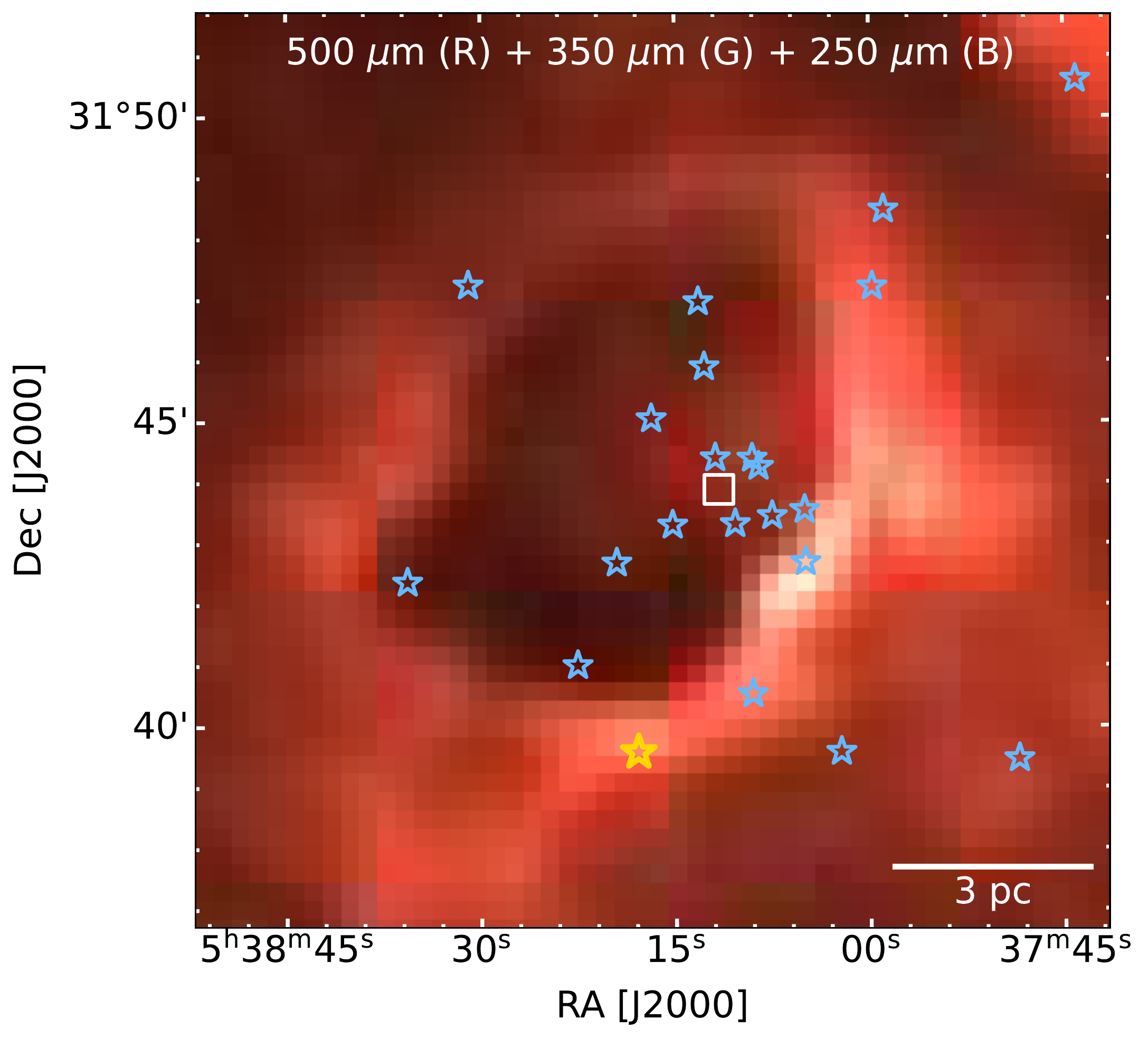}
    \includegraphics[width=0.45\textwidth]{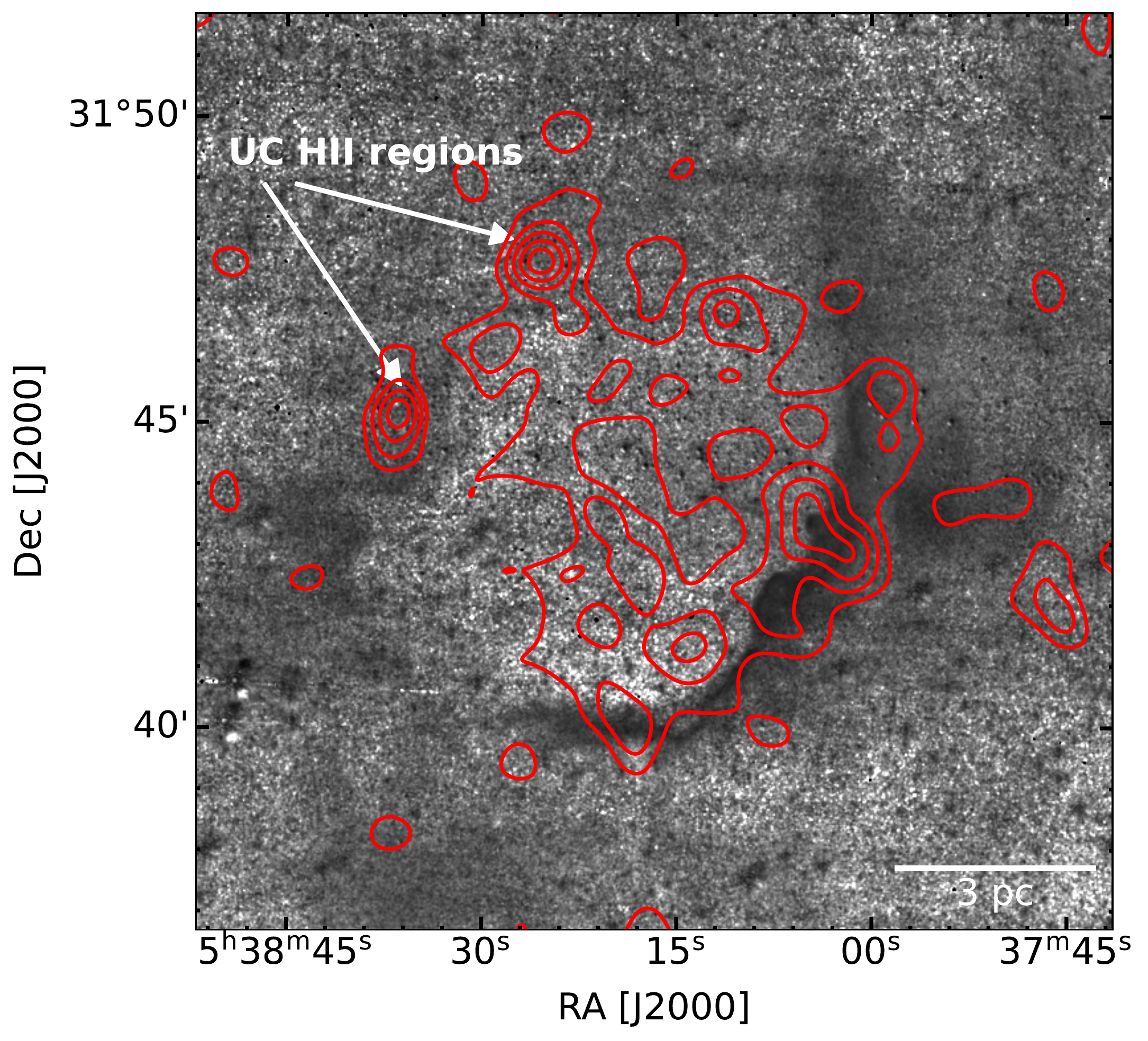}
    \includegraphics[width=0.45\textwidth]{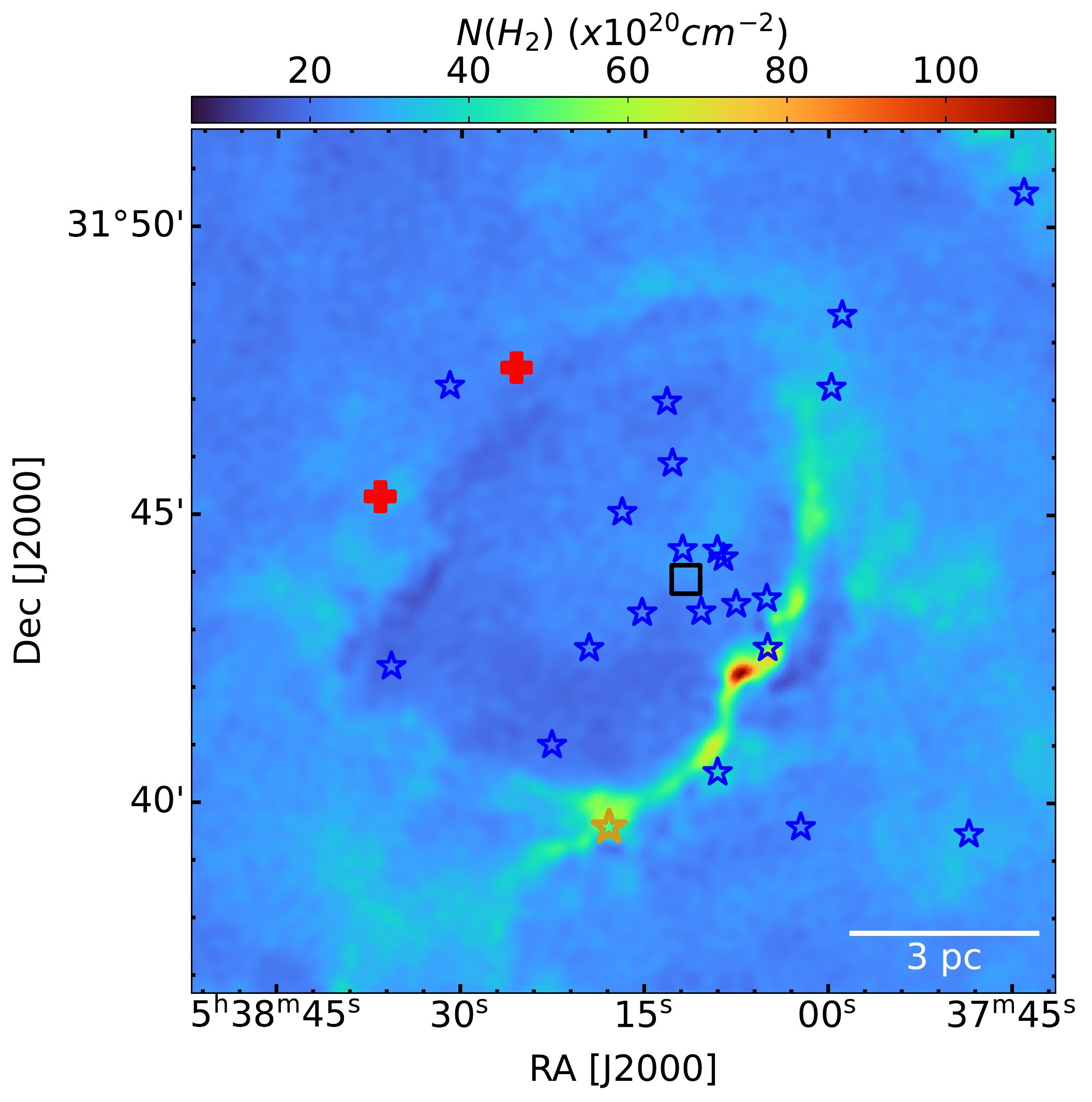}
    \includegraphics[width=0.45\textwidth]{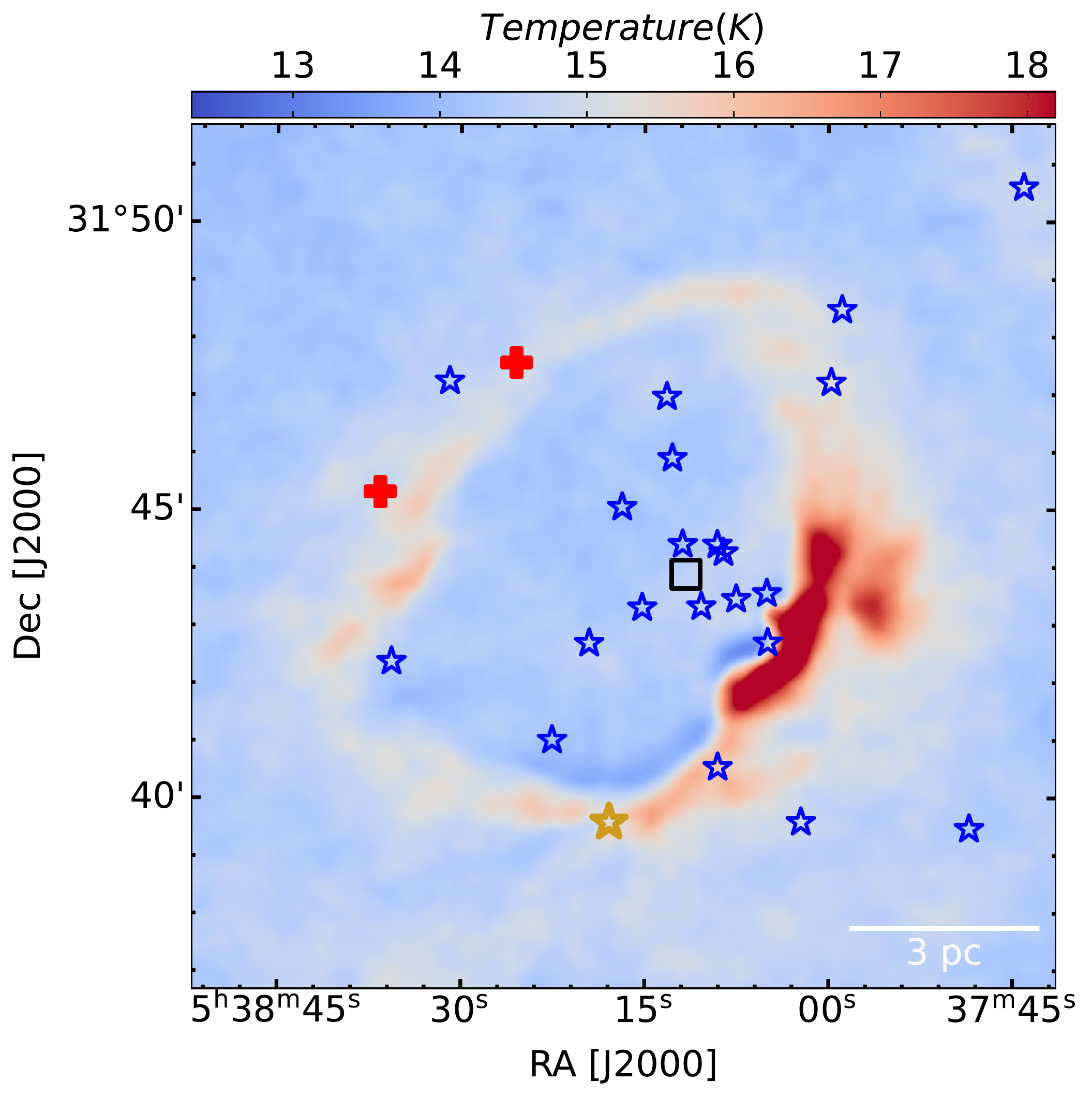}
    \caption{Upper left panel: Color-composite image (Red: \emph{Herschel} 500 $\mu$m; Green: \emph{Herschel} 350 $\mu$m, and \emph{Herschel} 250 $\mu$m) of $15^\prime \times15^\prime$ region overlaid with the locations of Class $\textsc{i}$ (yellow asterisk) and Class $\textsc{ii}$ (blue asterisk) YSOs. The location of identified massive star M1 is also marked with a white square. Upper Right panel: \emph{Spitzer} ratio map of 4.5 $\mu$m/3.6 $\mu$m emission, smoothed using Gaussian function. The ratio map is overlaid with the NVSS 1.4 GHz radio continuum contours (red). The lowest level to generate these contours is 0.6 mJy\,beam$^{-1}$ with a step size of 0.5 mJy\,beam$^{-1}$. Lower left panel: \emph{Herschel} column density map. Lower right panel: \emph{Herschel} temperature map. Both the temperature and column density maps are overlaid with the locations of Class $\textsc{i}$ (brown asterisk) and Class $\textsc{ii}$ (blue asterisk) YSOs along with the location of massive star `M1' (black square). The locations of UC regions are also marked with red `$+$'.}
    \label{fig:rgb_images}
\end{figure*}

The recently available archival wide-field IR, submm, and radio maps can give us a clear picture of the distribution of young stars, gas, dust, ionized gas, etc., which will help us to better understand the star formation scenario in the region \citep{2010A&A...523A...6D, 2017ApJ...834...22D}. In Figure \ref{fig:multiband}, we show the multi-wavelength picture of our selected region of the E70 bubble, starting from 1.25 $\mu$m up to 500 $\mu$m. The shorter wavelength mostly shows the photospheric emission from the stellar sources, and as we go towards the longer wavelength, we can see the prominent ring/structure of the E70 bubble made from gas and dust.

The upper left panel of Figure \ref{fig:rgb_images} depicts the color-composite image of E70 generated using the \emph{Herschel} 500 $\mu$m (red), 350 $\mu$m (green), and 250 $\mu$m (blue) band images. The \emph{Herschel} FIR images manifest a ring or a shell-like structure of the E70 bubble and the peak intensity is found at the arc of the bubble towards the west direction. The cold dust components are responsible for \emph{Herschel} FIR (160-500 $\mu$m) emission. The locations of YSOs and M1 are also shown on the image. Here it is worthwhile to note that a younger Class $\textsc{i}$ YSO is located on the arc-like structure and the other YSOs are located mostly inside the ring/shell-like structure.

The upper right panel of Figure \ref{fig:rgb_images} demonstrates the \emph{Spitzer} ratio map of 4.5 $\mu$m/3.6 $\mu$m emission smoothed using Gaussian function with two-pixel radius. Due to the same point spread function of 3.6 $\mu$m and 4.5 $\mu$m bands, these can be directly divided which permits us to remove the point-like sources along with continuum emission (cf. \citealt{2017ApJ...834...22D}). Some bright and dark regions are visible in this ratio map. The brighter region traces the 4.5 $\mu$m emission indicative of a prominent Br-$\alpha$ emission at 4.05 $\mu$m and a molecular hydrogen line emission ($\nu = 0-0 S(9)$) at 4.693 $\mu$m; on the other hand, darker region traces the 3.6 $\mu$m emission indicative of the presence of PAH emission at 3.3 $\mu$m and creates a PDR. This PDR might have been produced by the interaction of strong UV radiation from the massive star/s with the surrounding molecular cloud. This ratio map is also overlaid with the NVSS 1.4 GHz radio continuum emission shown by red contours. We see the diffuse radio emission enclosed in the MIR bubble/PDRs. This kind of radio morphology is a typical feature of the H\,{\sc ii} region/MIR bubble created by massive star/s \citep{2010A&A...518L.101Z,2014A&A...566A.122S}. On the boundary of the bubble, we have marked two separate radio peaks which might be created due to the effect of very young massive stars (ultra-compact (UC) H\,{\sc ii} region) formed on the shell of the bubble in case of collect and collapse scenario \citep{2007prpl.conf..181H,2014A&A...566A.122S}. The UC H\,{\sc ii} regions are of great interest morphologically as they provide hints about the state of the surrounding medium as soon as the massive star has formed. 

In the lower left panel of Figure \ref{fig:rgb_images}, we represent \emph{Herschel} column density map of the region to get a signpost of embedded structures. 
Locations of YSOs and `M1' are also marked in the image. The column density map also clearly shows the bubble structure with higher column density along the arc of the bubble. The western boundary which is nearer to `M1' seems to have more molecular material in comparison to other parts of the ring. The only Class $\textsc{i}$ YSO is also located on the ring at higher column density.

The lower right panel of Figure \ref{fig:rgb_images} represents the corresponding temperature map. The arc of the bubble showing high column density exhibits dust emission warmer (i.e., $T_d \sim$ 17-18 K) than the surroundings. The western arc of the ring seems to have the highest temperature which might be due to feedback from the massive star.


\begin{figure*}
\centering
\includegraphics[width=0.98\textwidth]{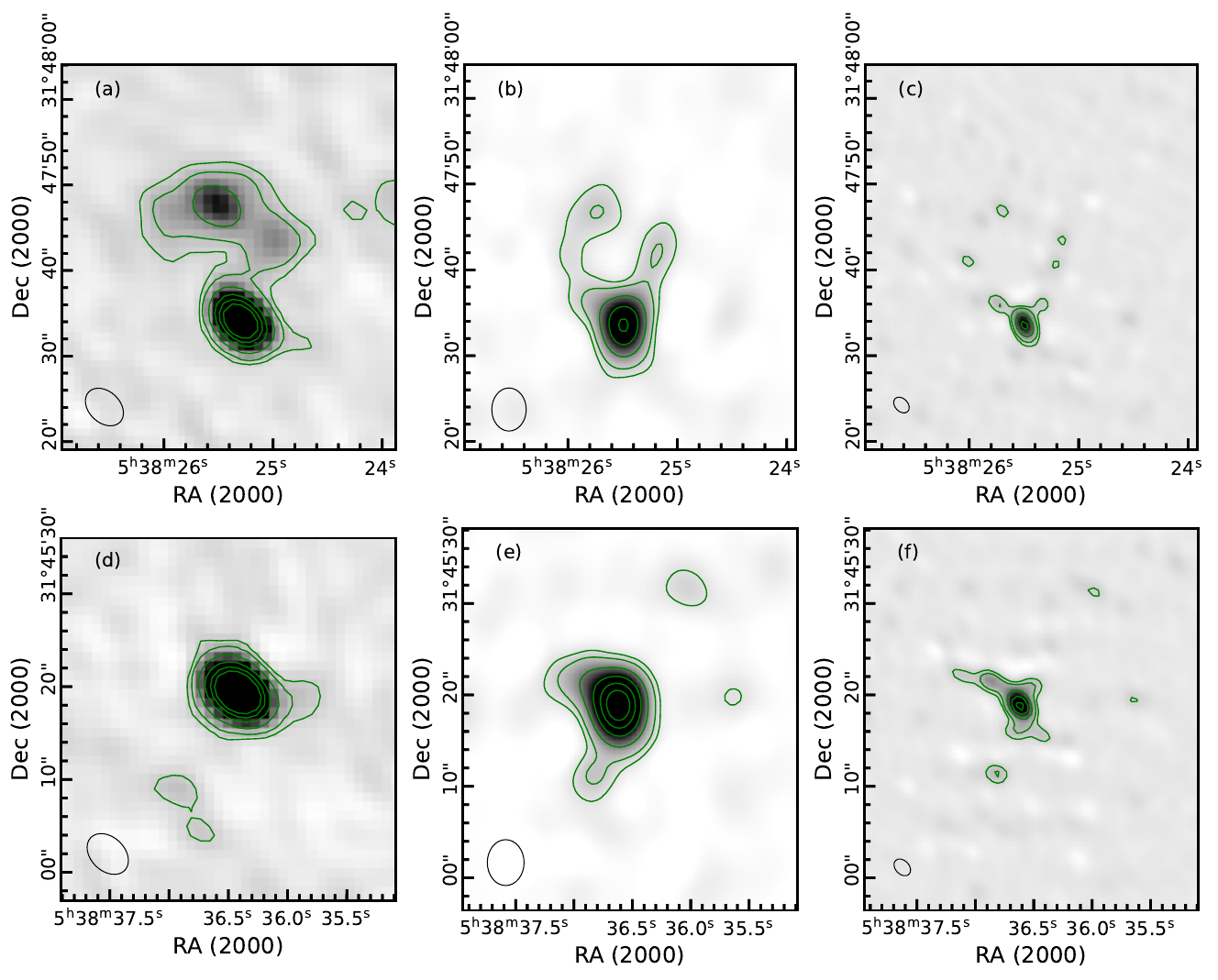}
\caption{
uGMRT radio maps for the Northern (upper row) and Southern H\,{\sc ii} (lower row)
regions.
(a) 610\,MHz image for the Northern UC H\,{\sc ii} region.
(b) 1400\,MHz lower resolution image for the Northern UC H\,{\sc ii} region.
(c) 1400\,MHz higher resolution image for the Northern UC H\,{\sc ii} region.
(d) 610\,MHz image for the Southern UC H\,{\sc ii} region.
(e) 1400\,MHz lower resolution image for the Southern UC H\,{\sc ii} region.
(f) 1400\,MHz higher resolution image for the Southern UC H\,{\sc ii} region.
The respective beams have been shown in the lower left corner (also see
Table \ref{table_GMRTObs}).
The contour levels for all the images are at 0.1, 0.2, 0.5, 1, 1.5,
and 2 mJy\,beam$^{-1}$.}
\label{fig_GMRTMaps}
\end{figure*}

\begin{figure}
\centering
\includegraphics[width=\linewidth]{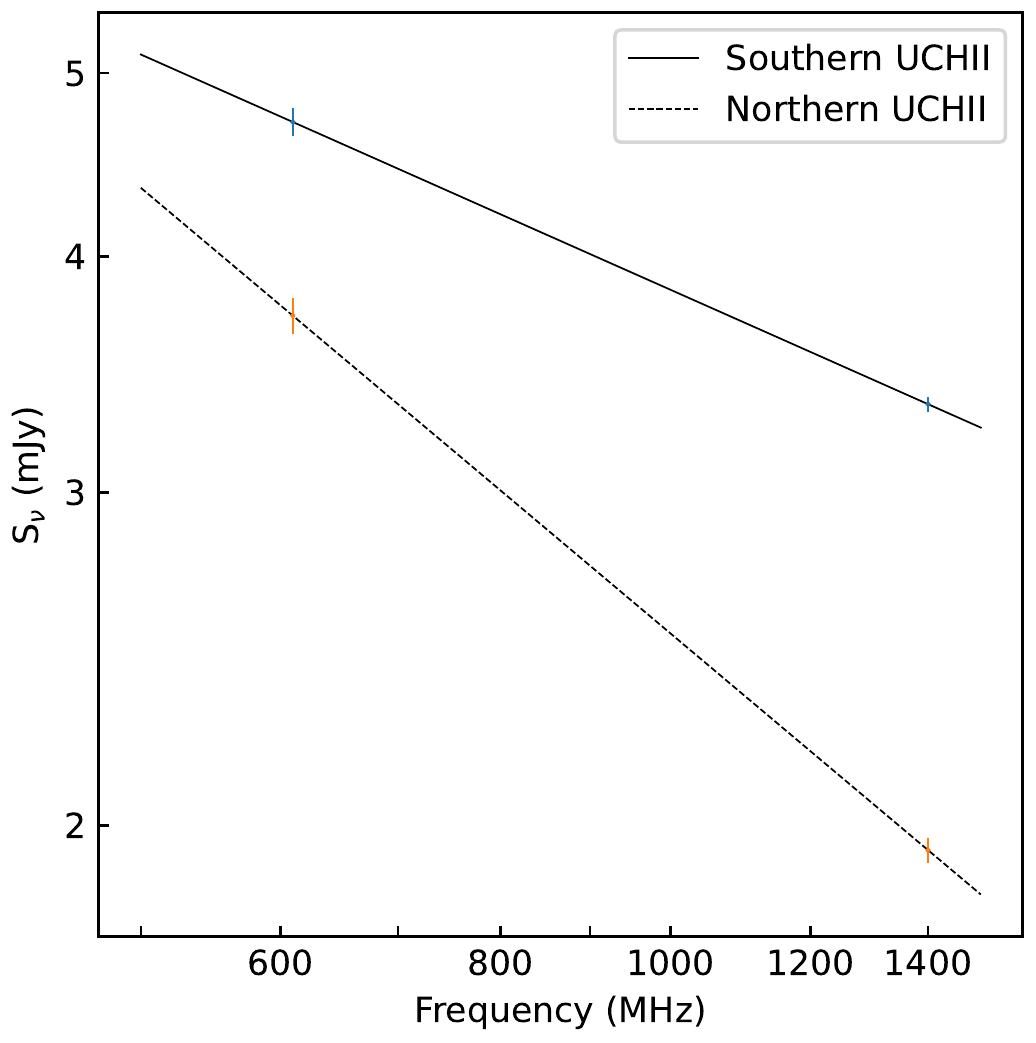}
\caption{
Log-log plot of the radio spectrum. The two data points (i.e. flux at 610 and 1400\,MHz at similar resolutions) for each region have been marked along with the respective error bars. The spectral indices ($\alpha$) for the two UC H\,{\sc ii} regions were calculated assuming S$_{\nu}$\,$\propto$\,$\nu^{\alpha}$.}
\label{fig_GMRT_SpectralIndex}
\end{figure}

\subsection{Radio Morphology: high-resolution uGMRT observations}
\label{section_radiomorphology}

We have seen the diffuse morphology of radio emission within the MIR bubble through NVSS contours. The same radio contour also hints towards two radio peaks on the edge of the eastern periphery of the bubble which might be due to the effect of very young and massive stars. Thus, to explore further, we have done a very high-resolution and sensitive observation of these regions through uGMRT. Figure \ref{fig_GMRTMaps} shows the morphology of the compact H\,{\sc ii} regions detected along the eastern periphery as seen at \emph{Herschel} FIR wavelengths (cf. upper left panel of Figure \ref{fig:rgb_images}) of the bubble. The first row (Figures \ref{fig_GMRTMaps}(a), (b), and (c)) and second row (Figures \ref{fig_GMRTMaps}(d), (e), and (f)) show the structure of the Northern and Southern compact H\,{\sc ii} regions, respectively. 
The AIPS task {\sc jmfit} was utilized to fit the emission regions with Gaussian's, yielding fitted parameters like the integrated intensity, peak intensity, and size of these region/s (cf. Table \ref{table_GMRTObs}). The relatively lower resolution images (Figures \ref{fig_GMRTMaps}(a), (b), (d), and (e)) for the two compact H\,{\sc ii} regions indicate the sizes to be of the order of $\sim$\,5\arcsec, (Table \ref{table_GMRTObs}) which translates to about 0.08 pc at a distance of 3.26 kpc, thus suggesting that the two regions are UC H\,{\sc ii} regions \citep{Kurtz_UCHIIRegions_2002}. While the Northern UC H\,{\sc ii} region shows an emission cavity to the immediate north of the main emission peak, the Southern UC H\,{\sc ii} region mostly displays a simple elliptical shape. The higher resolution 1400\,MHz images (Figures \ref{fig_GMRTMaps}(c) and (f)) do show a more detailed structure, as well as the size of main emission region to be of the order of 2\arcsec-3\arcsec\, ($\sim$0.03-0.05 pc) which would make the main emission clumps at least to be hypercompact H\,{\sc ii} regions.

Since the highest resolution 610\,MHz image had a beam size of 5\arcsec.08 $\times$ 3\arcsec.65, we also constructed a similar-resolution 1400\,MHz image (5\arcsec $\times$ 4\arcsec) by convolving the high-resolution image with this beam size, using the AIPS task {\sc convl}. The flux derived from similar-resolution images was used to obtain the spectral index.  For the main emission clumps for both the UC H\,{\sc ii} regions, we calculated the spectral index $\alpha$ assuming flux S$_{\nu}$\,$\propto$\,$\nu^{\alpha}$, where $\nu$ is the frequency, obtaining $\alpha$ to be $-$0.78 $\pm$ 0.03 and $-$0.41 $\pm$ 0.02 for the northern and southern UC H\,{\sc ii} regions, respectively. The fits are shown in Figure \ref{fig_GMRT_SpectralIndex}.

In H\,{\sc ii} regions, the radio spectrum due to thermal free-free emission \citep{CondonRansom_NRAOBook} is expected to have an optically thick region ($0< \alpha \lesssim2$) and an optically thin region ($\alpha \sim-$0.10). However, here $\alpha$ has a large negative value which suggests significant nonthermal synchrotron emission for the two UC H\,{\sc ii} regions. Such a mechanism in H\,{\sc ii} regions is thought to be associated with shocks \citep{DeBecker_PACWB_2013AA, Ainsworth_2014ApJ, Veena_IRAS17256_2016MNRAS, Dewangan_S305_2020ApJ}. Though  rigorous modeling of the thermal and nonthermal components is outside the scope of this paper, we do the following basic calculation similar to \citet{Dewangan_S305_2020ApJ} to estimate the spectral types associated with the two UC H\,{\sc ii} regions. 
The standard calculation of flux from an H\,{\sc ii} region assumes the emission to be due to the free-free emission (thermal bremsstrahlung) in the plasma \citep{CondonRansom_NRAOBook,Matsakis_1976AJ}. Hence one needs to obtain an estimate of the thermal contribution in the total flux.
According to the modelling  by \citet{Dewangan_S305_2020ApJ}, for a nonthermal spectral index ($\alpha_{\textrm{\sc{nt}}}$) of -0.8, they have calculated 76\% of the flux at 1280\,MHz (uGMRT L-band) as the contribution from thermal emission component.
According to their model, if $\alpha_{\textrm{\sc{nt}}}$ were to become steeper (i.e. more negative), the contribution of nonthermal emission component at a frequency will decrease and that of thermal emission component increase. The northern UCHII region has a similar spectral index (-0.78), and thus taking 76\% as contribution of thermal plasma at 1400\,MHz, the Lyman continuum flux 
was calculated using the following formula from \cite{Matsakis_1976AJ} : 
\begin{equation}
\small
 L_{c} = 7.5 \times 10^{46} \left(\frac{S}{Jy}\right) \left(\frac{D}{kpc}\right)^2 \left(\frac{\nu}{GHz}\right)^{0.1} \left(\frac{T_e}{10^4 K}\right)^{-0.45},
\end{equation}
where S is the flux density in Jy, D is the distance in kpc, $\nu$ is the frequency in GHz (1.4 GHz for our calculation), and T$_e$ is the electron temperature. Taking distance to be 3.26\,kpc and the electron temperature to be 10$^4$\,K \citep{CondonRansom_NRAOBook}, L$_c$ comes out to be $\sim$10$^{45.09}$
photons\,s$^{-1}$, which on comparison with the tabulated values of \citet[assuming ZAMS]{1973AJ.....78..929P} would suggest a spectral type of B1-B2 for the northern UCHII region. For the southern UCHII region, the spectral index is less steep (-0.41), and thus the contribution of thermal emission component will be lesser here. Taking a conservative value of 50\% returns a Lyman flux of $\sim$10$^{45.14}$ photons\,s$^{-1}$, and would suggest a B1-B2 type source (assuming ZAMS) for this region as well.


\begin{figure*}
\centering
\includegraphics[width=0.98\textwidth]{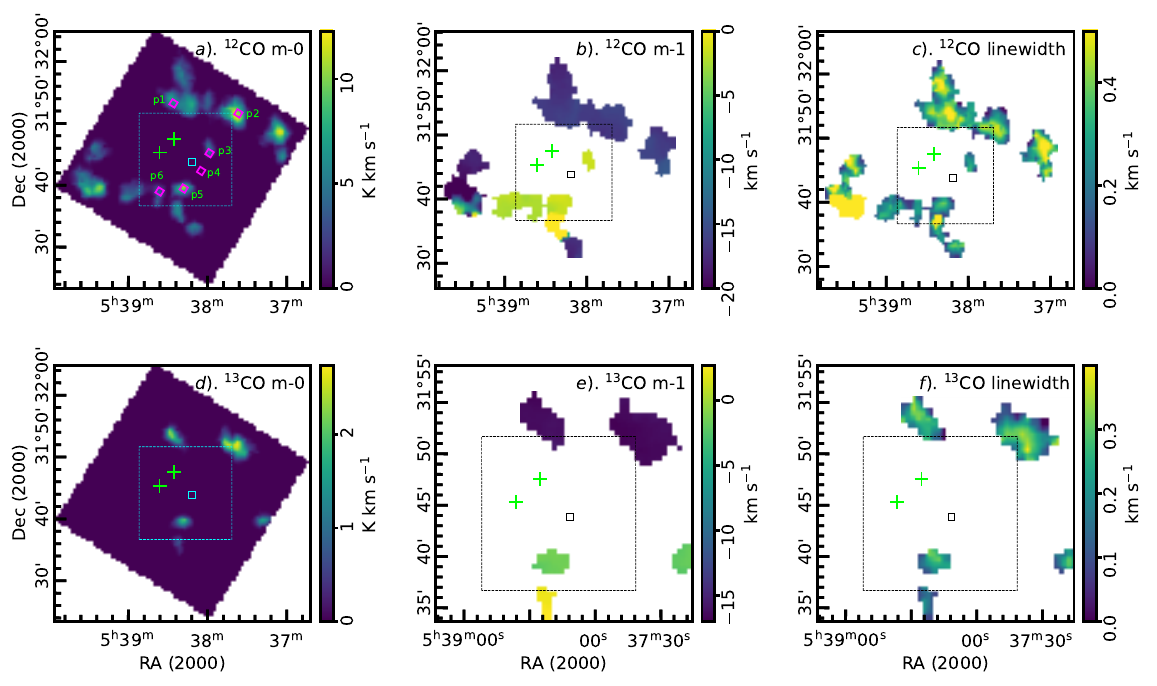}
\caption{
Column-wise :
moment-0 (Integrated intensity), moment-1 (Intensity-weighted velocity), and linewidth (Intensity-weighted dispersion) collapsed images for the two cubes -- $^{12}$CO$(J=1-0)$ and $^{13}$CO$(J=1-0)$ in the first and second rows, respectively. The emission regions seen in the $^{12}$CO and $^{13}$CO maps are those above 5$\sigma$ and 3$\sigma$, respectively ($\sigma$ being the rms noise of the respective spectral cubes). The green plus symbols and the square symbol mark the locations of UC H\,{\sc ii} regions and the massive star, respectively. The dashed box marks the FOV of Figure \ref{fig:rgb_images}. In (a), the labeled (p1-p6) magenta diamond symbols mark the positions where $^{12}$CO and/or $^{13}$CO spectra were extracted.
}
\label{fig_CO_m012}
\end{figure*}

\begin{figure*}
\centering
\includegraphics[width=0.98\textwidth]{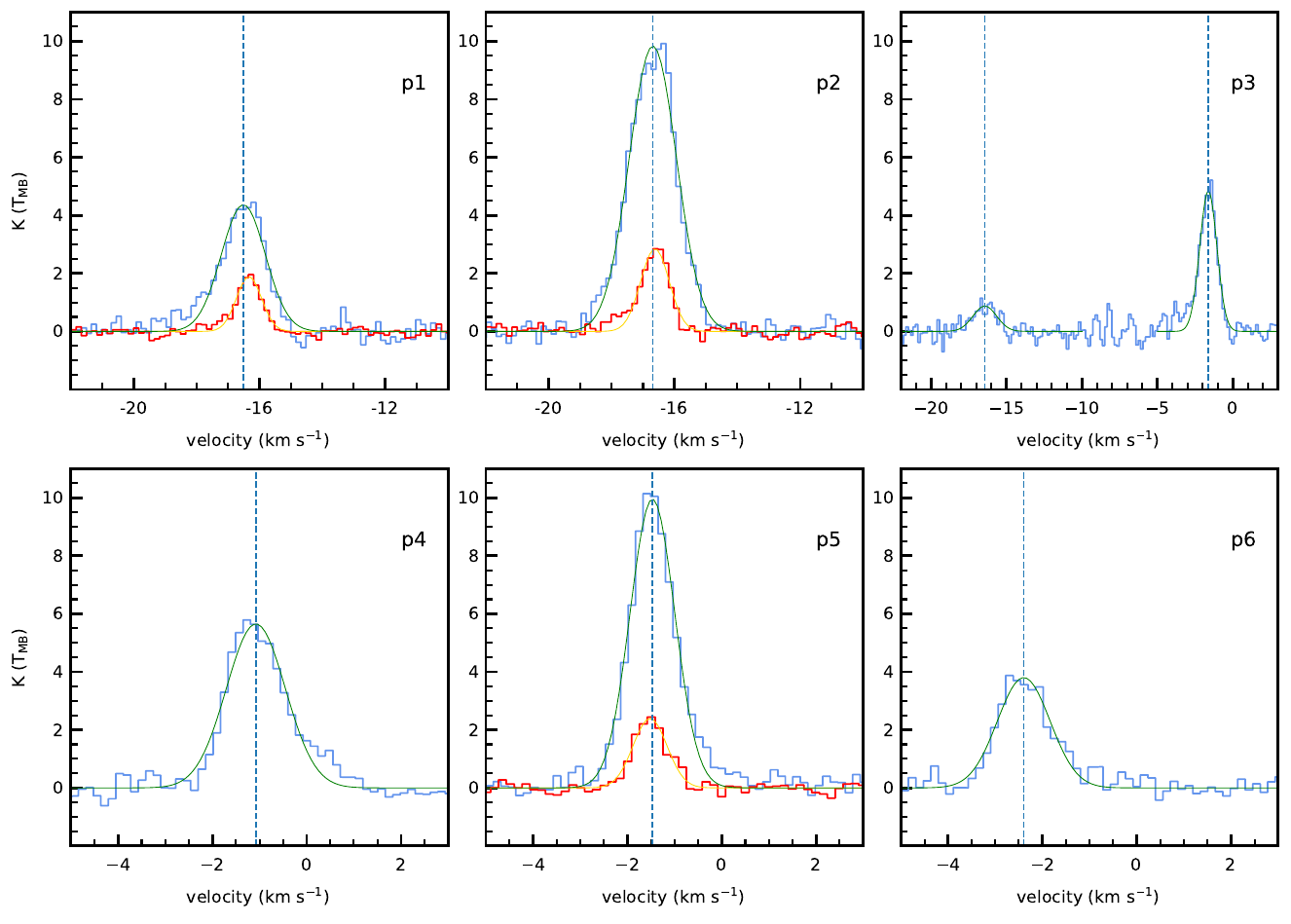}
\caption{
$^{12}$CO (J = 1-0) (blue) and $^{13}$CO (J = 1-0) (red) spectra at positions marked in Figure \ref{fig_CO_m012}. The Gaussian fits to the respective spectra are shown in green and yellow curves, with a blue dashed line marking the peak velocity of the $^{12}$CO Gaussian fit/s.
}
\label{fig_CO_Spectra}
\end{figure*}


\subsection{Molecular (CO) Morphology}
\label{section_MolecularMorphology}

\begin{table}
\caption{Parameters derived from $^{12/13}$CO spectra
         (Figure \ref{fig_CO_Spectra}) at the locations
         p1-p6 in Figure \ref{fig_CO_m012}(a).}
\label{table_COspectrum}
\begin{tabular}{lll}
\hline
Position  &       $v\,\pm\,\sigma$                          &       Amp.                \\
          &       (km\,\,s$^{-1}$)                             &       (K)                      \\
\hline
\multicolumn{3}{l}{$^{12}$CO} \\
\cmidrule(lr){1-1}
p1        &       $-$16.5            $\pm$   0.7              &       4.4                      \\
p2        &       $-$16.7            $\pm$   0.8              &       9.8                      \\
p3 (blue) &       $-$16.4            $\pm$   0.8              &       0.9                      \\
p3 (red)  &       $-$1.6             $\pm$   0.6              &       4.8                      \\
p4        &       $-$1.1             $\pm$   0.6              &       5.6                      \\
p5        &       $-$1.5             $\pm$   0.5              &       10.0                     \\
p6        &       $-$2.4             $\pm$   0.6              &       3.8                      \\
\hline
\multicolumn{3}{l}{$^{13}$CO} \\
\cmidrule(lr){1-1}
p1   &       $-$16.3            $\pm$   0.4              &       1.9                      \\
p2   &       $-$16.6            $\pm$   0.5              &       2.8                      \\
p5   &       $-$1.5             $\pm$   0.4              &       2.4                      \\
\hline
\end{tabular}
\end{table}

In this section, we examine the molecular morphology in a larger FOV ($\sim$30\arcmin$\times$30\arcmin) around the E70 bubble. To avoid confusing probable artefacts in the position-position-velocity (ppv) cubes with real physical features, we limit our analyses to only those regions which were detected above some threshold, taken to be 5$\sigma$ and 3$\sigma$ for the $^{12}$CO and $^{13}$CO cubes, respectively, with $\sigma$ being the rms noise level of the respective cubes. This was achieved using the following set of steps 
(see e.g., \citealt{2023ApJ...944..228M,2023JApA...44...34M}): identification of clumps in ppv space (above the requisite threshold) using the \emph{clumpfind} algorithm \citep{Williams_clumpfind_94} implemented via \textsc{starlink}'s \textsc{cupid} package \citep{Currie_starlink_2014, Berry_StarlinkCupid_2007}; masking those regions with no detection of clumps to create a masked cube; and collapsing the cubes to finally obtain moment-0 (m-0 or Integrated Intensity), moment-1 (m-1 or Intensity-weighted velocity), and linewidth (Intensity-weighted dispersion) maps. Figure \ref{fig_CO_m012} shows the m-0, m-1, and linewidth maps for the region in $^{12}$CO (J = 1-0) and $^{13}$CO (J = 1-0) transitions. We note that since for the C$^{18}$O (J = 1-0) transition, neither any features were visible in any of the channels nor was there any detection of clumps at 3$\sigma$ level, we restrict our further discussion to $^{12}$CO and $^{13}$CO cubes only.

As $^{12}$CO (J = 1-0) transition traces molecular gas structures of a typical density of $\sim$\,10$^2$\,cm$^{-3}$, while $^{13}$CO (J = 1-0) transition traces that for $\sim$\,10$^{3-4}$\,cm$^{-3}$ \citep{Su_MWISP_2019ApJS}, more detailed structures are seen in Figure \ref{fig_CO_m012}(a) as compared to only a few ($^{13}$CO) clumps in Figure \ref{fig_CO_m012}(d). The massive star (square symbol) and the two UC H\,{\sc ii} regions (plus symbols) have been marked in Figure \ref{fig_CO_m012}. The six positions marked with magenta boxes (labeled p1-p6) in Figure \ref{fig_CO_m012}(a) are locations where we extracted the molecular spectra.

The first thing we notice is that most of the $^{12}$CO emission are along the north and south edge of the image (Figure \ref{fig_CO_m012}(a)), with a narrow emission structure (along p3-p5) connecting the northern and southern emission structures. The massive star (marked with a square symbol) is almost at the center of these two northern and southern emission structures. In Figure \ref{fig_CO_m012}(b), which traces the velocity, we see a stark contrast between the two edges, with the northern structure having a blue-shifted velocity and the southern structure having a red-shifted velocity. To further examine the same, the extracted spectra at the positions p1-p6 (Figure \ref{fig_CO_m012}(a)) have been shown in Figure \ref{fig_CO_Spectra}. At position p4, while no clump was detected either in $^{12}$CO or $^{13}$CO at the requisite threshold, the spectrum was extracted as it lies near the massive star. The positions p1, p2, and p5 correspond to clumps that were detected in both the transitions (see Figure \ref{fig_CO_m012}(d)). The results of the Gaussian fits to these spectra are given in Table \ref{table_COspectrum}. It can be seen that the northern emission structure has a nearly constant mean velocity of $\sim-$16.5 \kms. As we move southwards, at position p3, the $-$16.5 \kms\, velocity component's strength is much reduced, and there is another component at $\sim-$1.6 \kms. Moving further southwards; along p4, p5, and p6; the mean velocity of the cloud is  ranging from $\sim-$1.1 to $-$2.4 \kms. At the eastern end of the image from p6 (see Figure \ref{fig_CO_m012}(b)), though the velocity seems to be at $\sim-$20 \kms \,in the m-1 map, it is likely some other cloud complex due to its distance from the bubble and very different velocity profile.


\subsection{Feedback Pressure exerted by the Massive Star}\label{sec:pressure}

Massive star affects its surroundings through feedback pressure which plays a crucial role in the self-regulation of star formation. This feedback pressure comprises of three components (see \citealt{2012ApJ...758L..28B} for details):
\begin{enumerate}
    \item Pressure exerted by H\,{\sc ii} region:
    \begin{equation}
        P_{HII} = \mu m_{H} c_{s}^2\, \left( \sqrt{\frac{3N_{uv}}{4\pi\,\alpha_{B}\, D_{s}^3}} \right),
    \end{equation}
    \item Radiation Pressure:
    \begin{equation}
        P_{rad} = \frac{L_{bol}}{4\pi c D_{s}^2}, 
    \end{equation}
    \\and\\
    \item Ram pressure exerted by stellar wind:
    \begin{equation}
        P_{wind} = \frac{\dot{M}_{w} V_{w}}{4 \pi D_{s}^2}.
    \end{equation}
\end{enumerate}

Here $\mu$ denotes the mean molecular weight of ionized gas ($=$ 0.678; \citealt{2009A&A...497..649B}), $m_H$ is the atomic mass of hydrogen, $c_s$ denotes the speed of sound in the photo-ionized region ($=11$ \kms; \citealt{2004fost.book.....S}), $N_{uv}$ is the Lyman continuum photons, $\alpha_{B}$ is radiative recombination coefficient ($= 2.6 \times (10^4 K / T_e)^{0.7}$ cm$^3$\,s$^{-1}$, \citealt{1997ApJ...489..284K}), $D_s$ is the projected distance between YSO core center and the massive star, $L_{bol}$ denotes the bolometric luminosity, $\dot{M}_{w}$ denotes mass-loss rate and $V_{w}$ denotes wind velocity of the ionizing source.

For O9V type star, $N_{uv} = 2.09 \times 10^{48}$ photons\,s$^{-1}$ \citep{1973AJ.....78..929P}, $\dot{M}_{w} = 1.20 \times 10^{-9}$ M$_{\odot}$ yr$^{-1}$ \citep{2009A&A...498..837M}, $V_w =$ 2200 \kms \citep{2017A&A...598A..56M} and
$L_{bol}$ = 79432.82 L$_{\odot}$ \citep{1973AJ.....78..929P}. We took $D_s =$ 0.46 pc as the projected distance between the cluster center and massive star `M1'. These values yield $P_{HII} = 1.12 \times 10^{-9}$, $P_{rad} = 4.01 \times 10^{-10}$, $P_{wind} = 6.58 \times 10^{-13}$ dynes\,cm$^{-2}$ and thus total pressure ($P_{HII} + P_{rad} + P_{wind}$) exerted by the massive star is $1.53 \times 10^{-9}$ dynes\,cm$^{-2}$.

\section{Discussion}\label{sec:discussion}

We have done a multi-wavelength analysis of the E70 bubble and its surrounding region to understand the star formation processes going on there. We have identified a cluster that is located within the E70 bubble. The size of this cluster is 1.7 pc and has a peak number density of 144.56 \,pc$^{-2}$.
The distance of the E70 cluster is estimated as 3.26 kpc through Gaia membership criteria and Bailer-Jones distance estimates \citep{2021AJ....161..147B}.
The obtained peak stellar number density is in good agreement with the peak number density (= 150 \,pc$^{-2}$) reported by \citep{2016AJ....151..126S} for the young clustering identified in the star-forming regions.
There are also several studies where a stellar clustering was found within a rim-like structure (or specifically a bubble), e.g., \citet{2017MNRAS.467.2943S} studied the NGC 7538 cluster region which was associated with MIR bubble, and star formation was triggered due to the feedback from the central massive star.

We have found a bright embedded star surrounded with warm gas/dust (as evident from 22 $\mu$m emission) inside the E70 bubble. The spectral type of this bright star is estimated as O9V using optical spectroscopy. This massive star is located within the E70 cluster boundary and the optical CMD confirms its membership to this cluster. As this is the brightest star in the E70 cluster, the upper age limit of the E70 cluster is estimated as $\sim$ 8.1 Myr.
This cluster also seems to host other massive stars having spectral type B1-B2 (refer to the right panel of Figure \ref{fig:optical_tcd}).  The abundance of the massive stars in the E70 cluster is also confirmed by the estimated MF slope value  ($\Gamma\simeq-$0.92), which is a bit shallower than the \citet{1955ApJ...121..161S} value, i.e.,  $\Gamma=-$1.35.
Similar slopes were also found in case of Sh2-301  \citep[$\Gamma=-$0.85 in the mass range $0.4<$ M/M$_\odot<7$,][]{2022ApJ...926...25P} and NGC 6910 \citep[$\Gamma=-$0.74 in the mass range $0.8<$ M/M$_\odot<25$,][]{2020ApJ...896...29K} which are massive star formation sites.

The massive star/s has/have a huge effect on its/their surroundings through their high energy UV radiations or stellar winds. These can trigger the formation of stars in two ways, either by compressing the preexisting dense molecular clump, known as ``radiation driven implosion'' (\citealt{1982ApJ...260..183S,1994A&A...289..559L}) or by sweeping out the adjacent molecular gas into the dense shell which fragments into prestellar cores afterward, known as ``collect and collapse process'' (\citealt{1977ApJ...214..725E,1998ASPC..148..150E}).
 The latter process has caught the eye of many astronomers since it behaves as a precursor for the formation of massive star/s or cluster/s. Various attempts have been made to understand the existence of the collect and collapse process (e.g., \citealt{2006A&A...446..171Z,2008A&A...482..585D,2010A&A...518L..81Z,2011A&A...527A..62B,2015ApJ...798...30L,2017A&A...606A...8D,2020ApJ...897...74Z,2023RAA....23a5011Z}). A ring (or arc) of gas and dust around the central cluster region containing massive stars, and the distribution of prestellar cores or young massive YSOs consistently spaced around the H\,{\sc ii} region, are one of the most important signatures for the collect and collapse scenario. We have also found  similar morphology, through the distribution of PDRs, warm and cold gas, dust/ionized gas, molecular condensation, etc.,  in our selected region. E70 demonstrates a ring/shell-like structure of gas and dust surrounding a massive star, as evident from the \emph{Herschel} maps (both intensity maps at different MIR bands and column density map). The temperature of the ring/shell is also a bit higher than the nearby region and has a distribution of PDRs especially near the massive star. The \emph{Spitzer} ratio map (4.5 $\mu$m/3.6 $\mu$m) also clearly shows the distribution of arc-shaped shocked structures having the distribution of PDRs. Inside the ring/shell, there is a distribution of diffuse radio emission which is indicative of the ionized gas in the region.

These morphological features confirm the strong feedback from the massive star `M1' to its surroundings. There are several observational evidences in the literature confirming that the strong feedback from the massive star/s can trigger star formation in their surrounding, e.g., \citet{2003A&A...408L..25D,2010A&A...518L..81Z,2011A&A...527A..62B,2020ApJ...905...61P} and \citet{2022ApJ...926...25P} etc. 
We have identified a YSO core (age $<$ 3 Myr) near the massive star `M1' (age $\sim$ 8.1 Myr) which might have formed due to the influence of the massive star itself. We have estimated the total pressure exerted by the massive star `M1' at the YSO core center as $1.53 \times 10^{-9}$ dynes\,cm$^{-2}$. For a typical molecular cloud, the internal pressure is of the order of $\sim 10^{-11}$ - $10^{-12}$ dynes\,cm$^{-2}$ for particle density $\sim 10^{3}$ - $10^{4}$ cm$^{-3}$ at a temperature of $\sim20$ K (cf. Table 2.3 of \citealt{1980pim..book.....D}). Thus, the feedback from M1 can in fact collapse the surrounding molecular cloud to create a new generation of stars. We have also found very young YSOs (Class $\textsc{i}$, age $<$ 0.5 Mys, \citealt{Evans_2009}) and two UC H\,{\sc ii} region (age $<$ 0.1 Myr, \citealt{1989ApJS...69..831W,1994ApJS...91..659K,2022A&A...666A..31M}) in the shell/ring around the massive star.
The diffuse population of the YSOs (age $<$ 3 Myr) outside the E70 cluster might have formed independently as it is not feasible to drift out a few pc away from the cluster boundary in the short time span of their formation.
Thus, there seems to be an age gradient around the massive stars `M', as the age of M1 is $\sim$ 8.1 Myr, and then there is a distribution of $\leq$ 3 Myr old YSOs within the E70 bubble, and then, the youngest population are located on the ring/shell of E70 bubble itself having age $<$ 0.5 Myr.

Finally, an age gradient, pressure calculations, a ring/shell of gas and dust, location of PAHs/PDRs on the ring/shell, ionized gas within the ring/shell, temperature/column density maps, location of very young Class $\textsc{i}$ YSO and UC H\,{\sc ii} region on the ring/shell, etc., all point toward positive feedback from a massive star `M1' and the collect and collapse scenario might be a possible model responsible for the formation of the youngest population of stars in the E70 bubble.  

\begin{figure}
\centering
\includegraphics[width=0.48\textwidth]{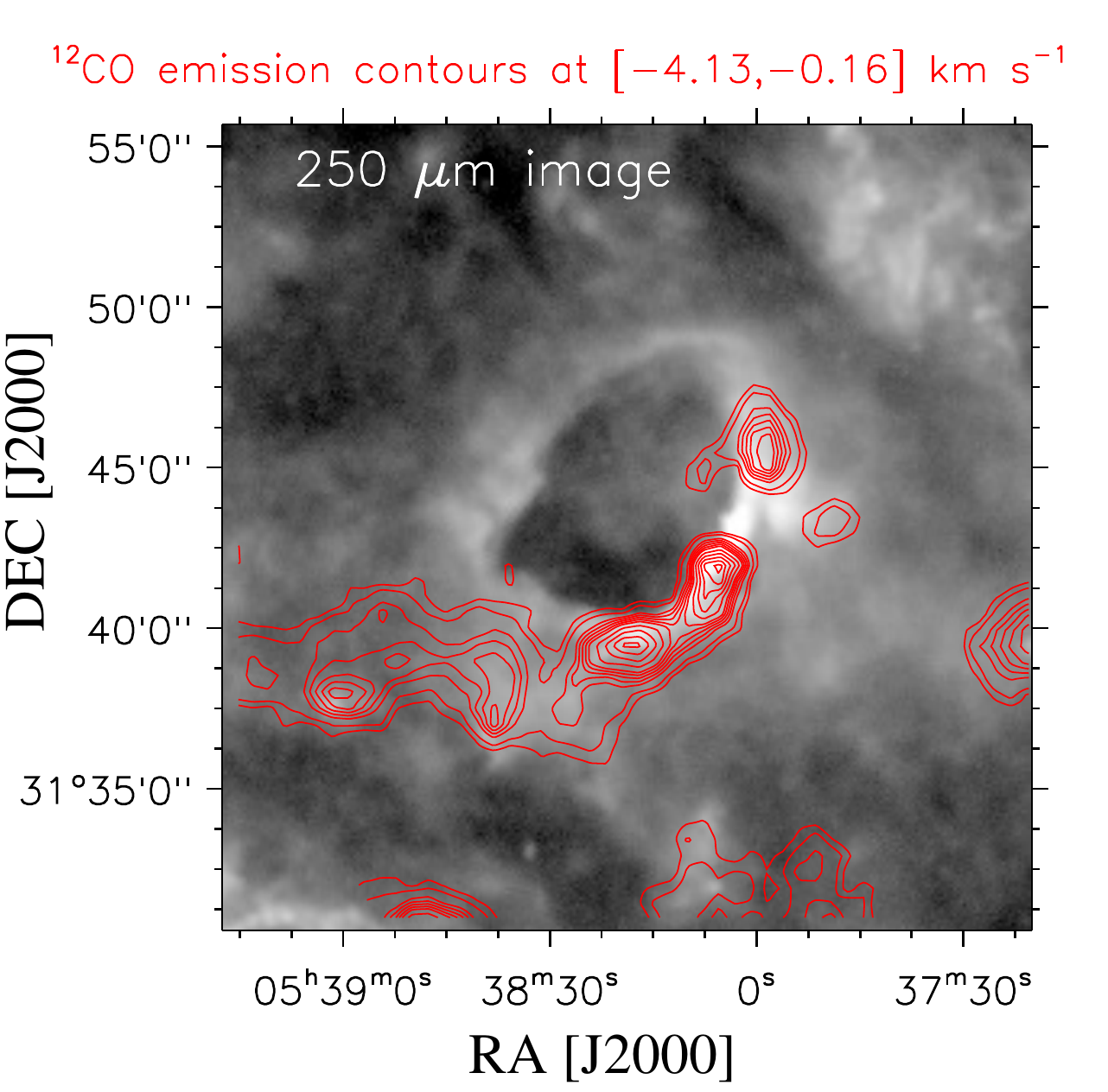}
\caption{$^{12}$CO (J = 1-0) emission contours (red curves) in the velocity range -4.13 to -0.16 km s$^{-1}$ overlaid on the \emph{Herschel} 250 $\mu$m image.}
\label{12co-wise}
\end{figure}

We have also studied the molecular morphology around the E70 bubble and it  seems that this region has a low-density molecular structures encompassing the bubble. In velocity space, the molecular clouds seems to have two distinct  velocity components, one  ($\sim-16.6$ \kms) in the north and another ($\sim-1.5$ \kms) in the southern direction from the massive star `M1'. 
Out of the identified molecular clumps (p1-p6), the ring/shell  of gas and dust  of E70 bubble seems to be associated with p3-p6 having a peak-velocity of $-$2.4 to $-$1.1 \kms (see also Table 3 and Figure \ref{12co-wise}).
As we go further away down south direction, there are more molecular condensation in the same velocity range and are probably the sites of further star formation.
This distribution of molecular clumps along the swept-up ring/shell of gas and dust further confirms our conclusion on the positive feedback of the massive star/s in the region.

\section{Conclusion}\label{sec:conclusion}

We present a multi-wavelength analysis of a Galactic MIR bubble `E70' using the data taken from various telescopes. The E70 bubble is an interesting region showing many signatures of active star formation, thus, our approach was to understand the star formation processes going on in this region. The following conclusions are made from our study:

\begin{enumerate}
      
	\item
	We have identified a stellar clustering inside the MIR bubble `E70' having almost circular morphology and small size (radius = 1.7 pc). We have also identified 29 members of this cluster using the Gaia DR3 PM data. Using the optical CMD, we have found a few probable massive stars which are located inside the E70 cluster boundary.

\item
	We have estimated the MF slope as $\Gamma\simeq-$0.92, which is shallower than the \citet{1955ApJ...121..161S} value, i.e.,  $\Gamma=-$1.35. This suggests the excess number of massive stars inside the E70 cluster in comparison to our solar neighborhood.

	\item
	We have done optical spectroscopy of the brightest member `M1' of the E70 cluster and estimated the spectral type as O9V. Thus, we have put an upper age limit to the E70 cluster as $\sim$ 8.1 Myr.

	\item
	We have estimated the distance of the E70 cluster as $3.26 \pm 0.45$ kpc and the minimum extinction value as $A_V$ = $2.64 \pm 0.20$ mag using the distribution of E70 cluster stars in the optical CMD and TCD.
      
	\item
	We have also found 22 stellar sources showing excess IR emission (YSOs with circumstellar discs around them), mostly located inside the E70 bubble. Most of these YSOs are found to be part of an embedded YSOs core which seems to be associated with the E70 cluster using the MST analysis. Since we have found both Class $\textsc{i}$ (age $<$ 0.5 Myr) and Class $\textsc{ii}$ (age $<$ 3 Myr) YSOs, we can safely conclude that the star formation is still going on in this region.

	\item
	Using the \emph{Herschel} MIR/FIR maps, we have identified the circular ring/shell-like structure of gas and dust around the massive star `M1'. The diffuse radio emission is found inside this bubble. The \emph{Spitzer} ratio map also hints towards the interaction of massive stars with the surrounding gas and dust.  The arc near the massive stars clearly shows the distribution of PDRs. The youngest Class $\textsc{i}$ YSOs is found to be located in the southern direction on the rim/arc of this bubble.

	\item
	Using the high-resolution uGMRT radio data, we have identified two UC H\,{\sc ii} regions located on the rim of the E70 bubble. The radio spectrum of these UC H\,{\sc ii} regions suggests both thermal and non-thermal components in the radio emission. The spectral type of the stars that probably are generating the UC H\,{\sc ii} regions is estimated as B1-B2.

    \item
     We have found that the total pressure exerted by the massive star `M1' ( $1.53 \times 10^{-9}$ dynes\,cm$^{-2}$) at the center of YSOs core is much higher than the typical internal pressure of a typical molecular cloud ($\sim 10^{-11}$-$10^{-12}$ dynes\,cm$^{-2}$).

    \item
     A ring/shell of gas and dust surrounding a massive star, location of PAHs/PDRs/very young YSO/UC H\,{\sc ii} region on the ring/shell, the age/pressure calculations, ionized gas within the ring/shell, temperature/column density maps, etc., all point toward a positive feedback from the massive star ‘M1’ and the collect and collapse scenario might be a possible model responsible for the formation of the youngest population of stars at the rim of the E70 bubble.

    \item
    We have found low-density molecular clumps having a paek-velocity of $-2.4$ to $-1.1$ \kms\, associated with the ring/shell of dust and gas of E70 bubble.
    It is possible that the massive star has swept up material to form a ring of gas and dust where a new generation of very young stars have formed.

\end{enumerate}

\begin{acknowledgments}
\section*{Acknowledgements}
We thank the anonymous referee for constructive and valuable comments that greatly improved the overall quality of the paper. The observations reported in this paper were obtained by using the 1.3m DFOT and 3.6m DOT telescopes at ARIES, Nainital, India and the 2m HCT at IAO, Hanle, the High Altitude Station of Indian Institute of Astrophysics, Bangalore, India. We thank the staff of the GMRT that made these observations possible. GMRT is run by the National Centre for Radio Astrophysics of the Tata Institute of Fundamental Research.
This research made use of the data from the Milky Way Imaging Scroll Painting (MWISP) project, which is a multi-line survey in 12CO/13CO/C18O along the northern galactic plane with PMO-13.7\,m telescope. We are grateful to all the members of the MWISP working group, particularly the staff members at PMO-13.7\,m telescope, for their long-term support. MWISP was sponsored by National Key R\&D Program of China with grant 2017YFA0402701 and by CAS Key Research Program of Frontier Sciences with grant QYZDJ-SSW-SLH047.
This publication makes use of data from the Two Micron All Sky Survey, which is a joint project of the University of Massachusetts and the Infrared Processing and Analysis Center/California Institute of Technology, funded by the National Aeronautics and Space Administration and the National Science Foundation. This work is based on observations made with the \emph{Spitzer} Space Telescope, which is operated by the Jet Propulsion Laboratory, California Institute of Technology under a contract with the National Aeronautics and Space Administration. This publication makes use of data products from the Wide-field Infrared Survey Explorer, which is a joint project of the University of California, Los Angeles, and the Jet Propulsion Laboratory/California Institute of Technology, funded by the National Aeronautics and Space Administration. DKO acknowledges the support of the Department of Atomic Energy, Government of India, under Project Identification No. RTI 4002. AV acknowledges the financial support of DST-INSPIRE (No.$\colon$ DST/INSPIRE Fellowship/2019/IF190550).
\end{acknowledgments}

\appendix

\section{Identification of YSOs in the Region}\label{app:yso_identification}

\begin{figure*}[!h]
\centering
    \includegraphics[width=0.3\textwidth]{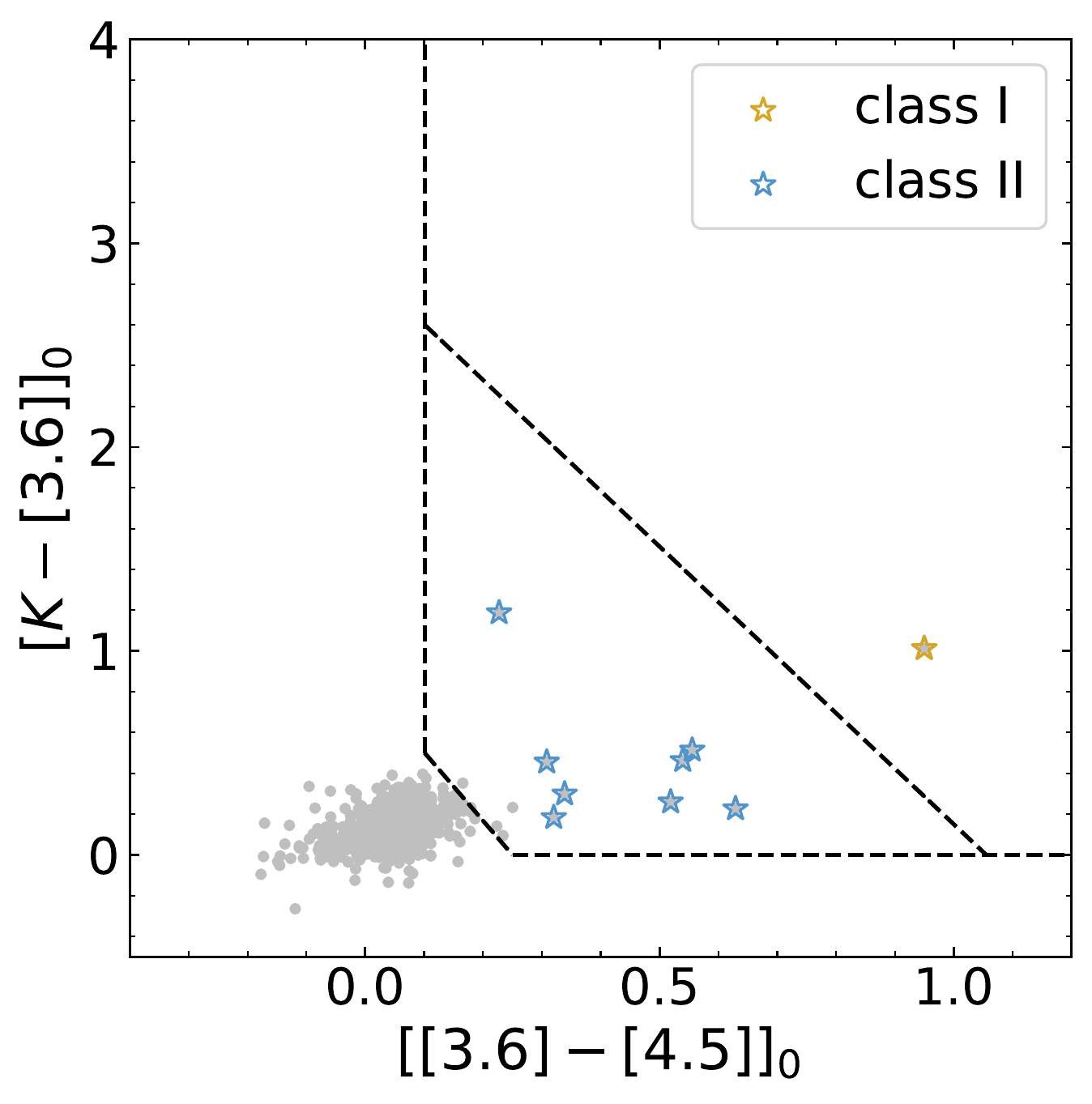}
    \includegraphics[width=0.33\textwidth]{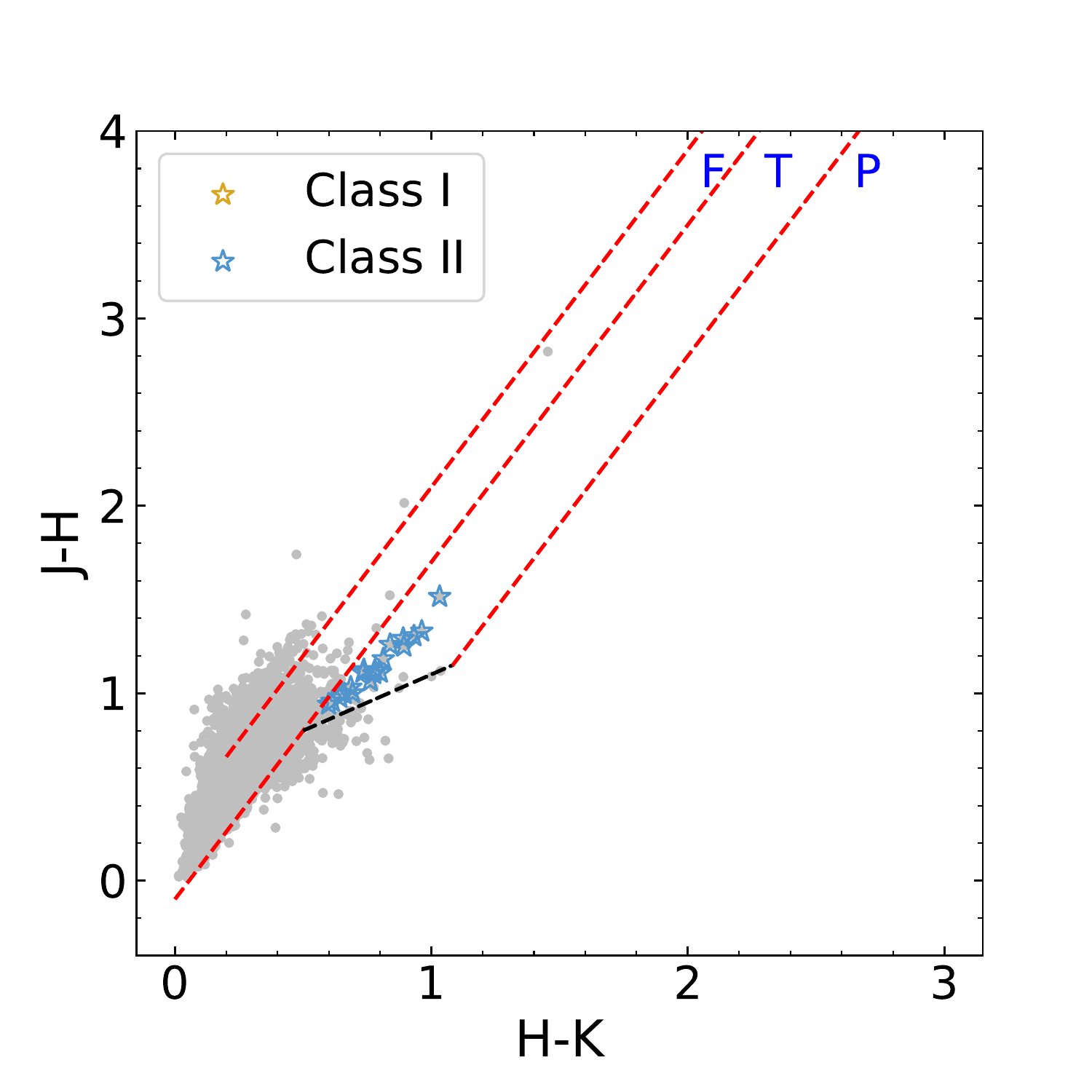}
    \includegraphics[width=0.3\textwidth]{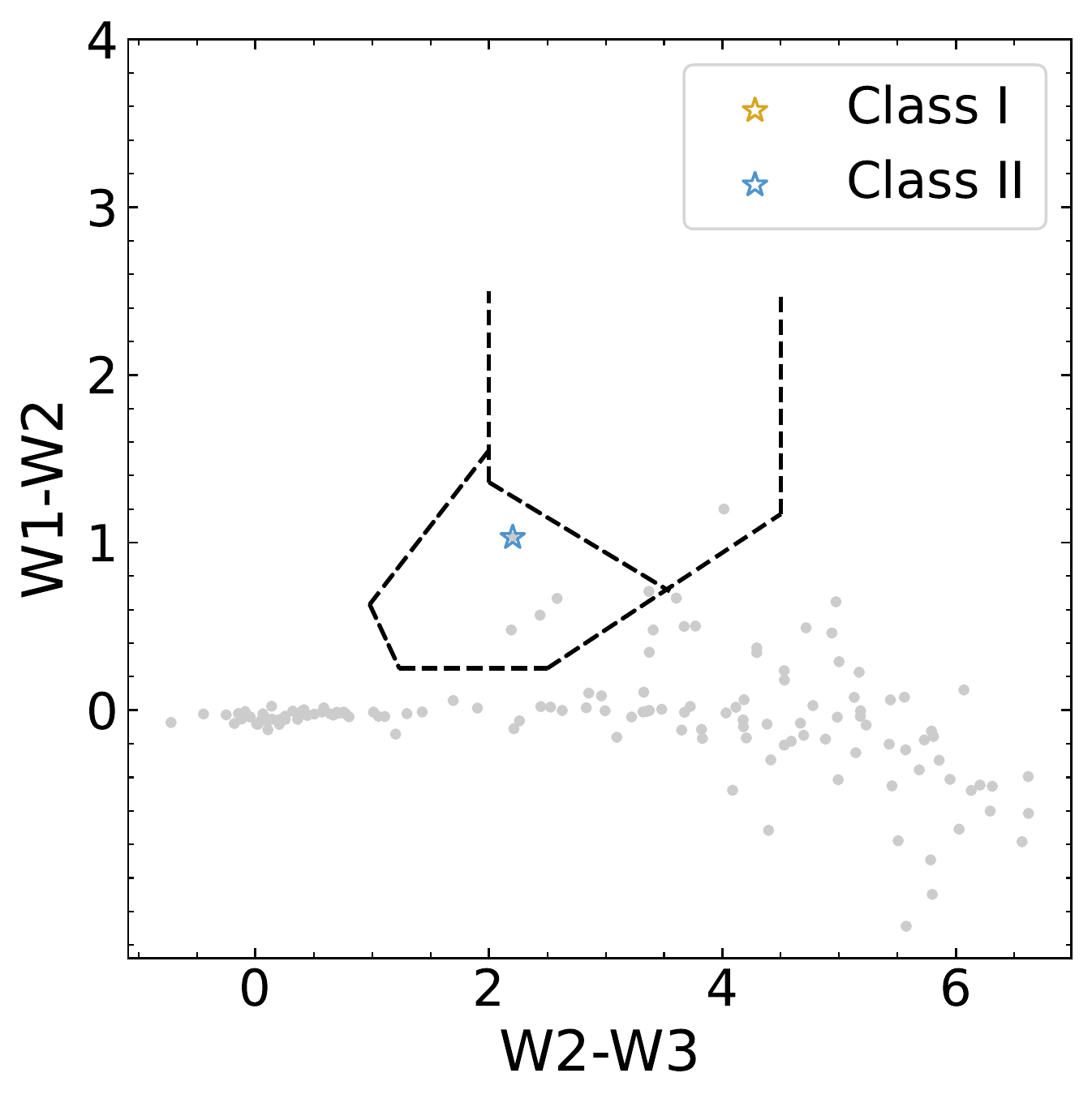}
    \caption{TCD for the YSOs identified within $15^\prime \times15^\prime$ region. Left panel: [K - [3.6]]$_0$ vs. [[3.6] - [4.5]]$_0$ TCD, classification of YSOs is based on the scheme given by \citet{2009ApJS..184...18G}. Middle panel: $(J-H)$ vs. $(H-K)$ CMD, classification is based on the scheme given by \citet{2004ApJ...608..797O}. The parallel dashed lines (red) represent the reddening lines drawn from the tip (spectral type M4) of the giant branch (left red line), from the base (spectral type A0) of the MS branch (middle red line) and from the tip of intrinsic CTTS line (right red line). Right Panel: $[W1-W2]$ vs. $[W2-W3]$ TCD, classification of YSOs is based on the scheme given by \citet{2014ApJ...791..131K}. Yellow and blue asterisks represent Class $\textsc{i}$ and Class $\textsc{ii}$ YSOs, respectively.}
    \label{fig:yso_classification}
\end{figure*}

YSOs are identified/classified on the basis of their excess IR emission. We have used the \emph{GLIMPSE360} catalog of \emph{Spitzer Science Centre} (SSC) by applying the updated classification scheme given by \citet{2009ApJS..184...18G} for the identification of YSOs. The [K - [3.6]]$_0$ versus [[3.6] - [4.5]]$_0$ two-color diagram (TCD; see left panel of Figure \ref{fig:yso_classification}) gives a total of 1 Class $\textsc{i}$ and 9 Class $\textsc{ii}$ YSOs in our region.

We have also used \emph{UKIDSS} data along with \emph{2MASS} data by choosing brighter sources, having \emph{J} magnitude $\leq$13 from \emph{2MASS} and the fainter sources, having \emph{J} magnitude $>$13 from \emph{UKIDSS}. In such a way we have created an NIR catalog. We have used the scheme given by \citet{2004ApJ...608..797O}. The (J-H) versus (H-K) TCD is plotted in the middle panel of Figure \ref{fig:yso_classification}. The three parallel lines are drawn from the tip of the giant branch, base of the main sequence branch and the tip of intrinsic Classical T Tauri stars (CTTS, these are basically reddening vectors. The sources falling in "F" region are either Class $\textsc{iii}$ sources or the field stars; in "T" region either Class $\textsc{ii}$ YSOs or CTSS ; and in "P" region are classified as Class $\textsc{i}$ YSOs. Using the scheme, we have identified a total of 15 Class $\textsc{ii}$ YSOs. The sources having their counterparts in \emph{Spitzer}, have been removed.

Furthermore, We have used the MIR data of \emph{ALLWISE} catalog of \emph{WISE} and followed the scheme given in \citet{2014ApJ...791..131K} for their classification. In this scheme, a selection criteria is applied on all four \emph{WISE} bands to get good and quality \emph{WISE} data then the extra galactic contaminants such as; AGNs, star-forming galaxies, transition disks are removed by applying the selection criteria on that \emph{WISE} data of the target. We have found out one Class $\textsc{ii}$ YSOs by using ([3.4] - [4.6]) versus ([4.6] - [12]) TCD (see right panel of Figure \ref{fig:yso_classification}).  
In total, we have obtained 22 YSOs showing excess IR emission.

\section{Membership Probability in the Cluster}\label{app:membership_prob}

\begin{figure*}[]
\begin{minipage}{0.5\columnwidth}
\includegraphics[width=\textwidth]{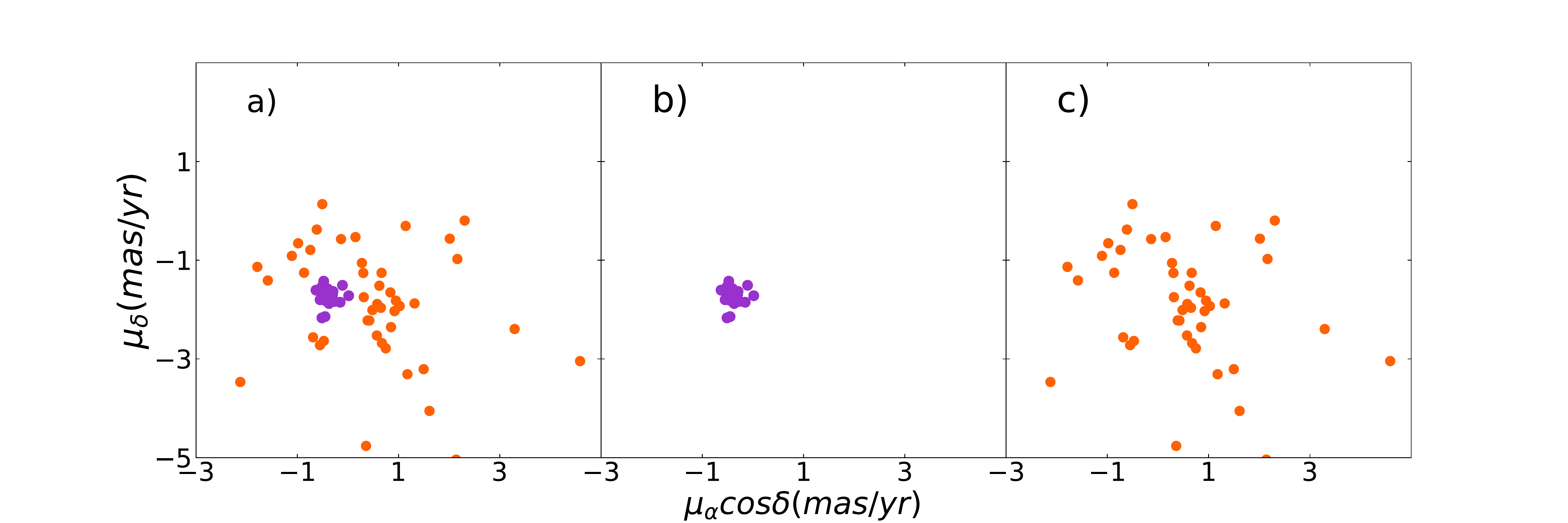}
\\[2mm]
\includegraphics[width=\textwidth]{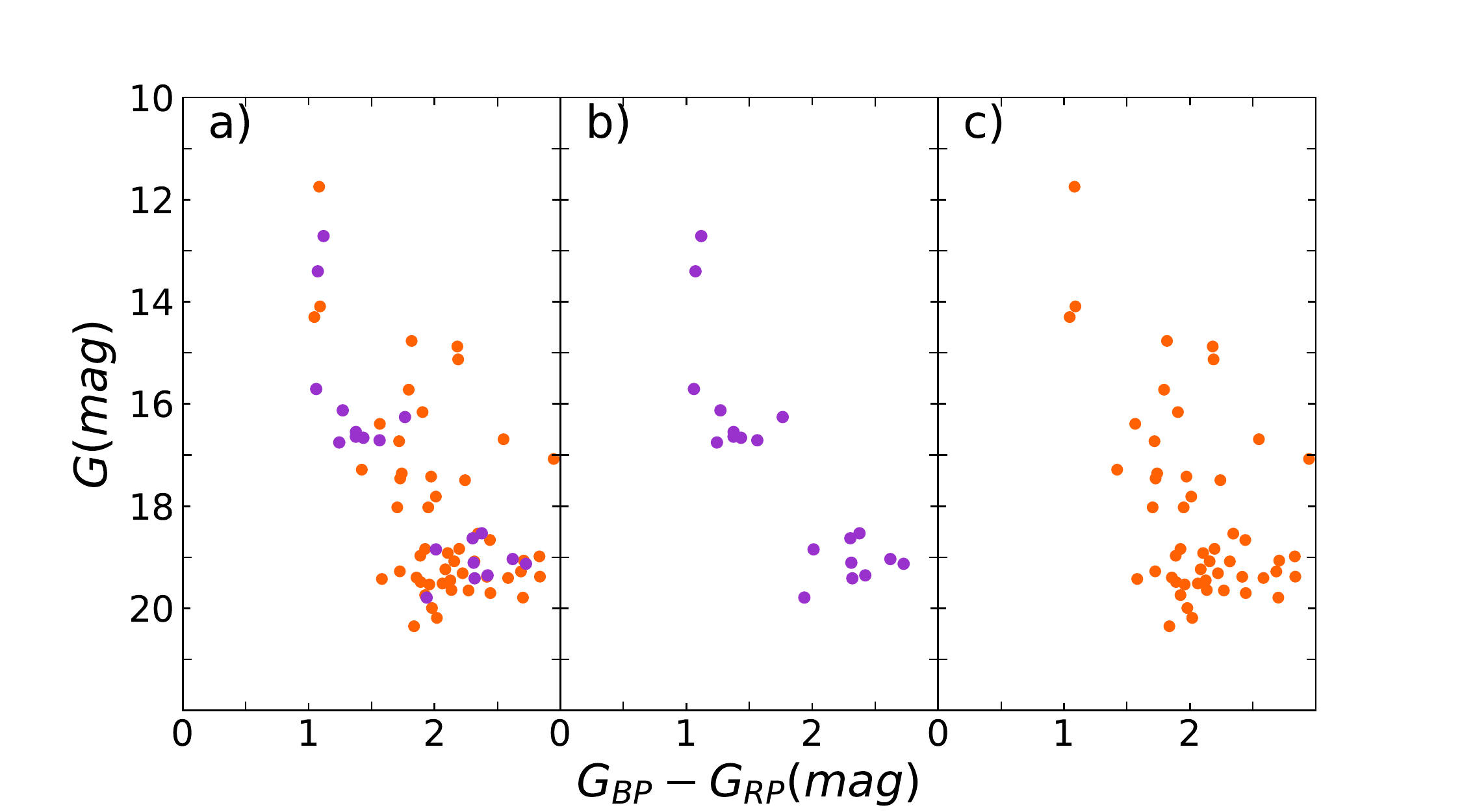}
\end{minipage}
\begin{minipage}{0.5\columnwidth}
\includegraphics[width=\textwidth]{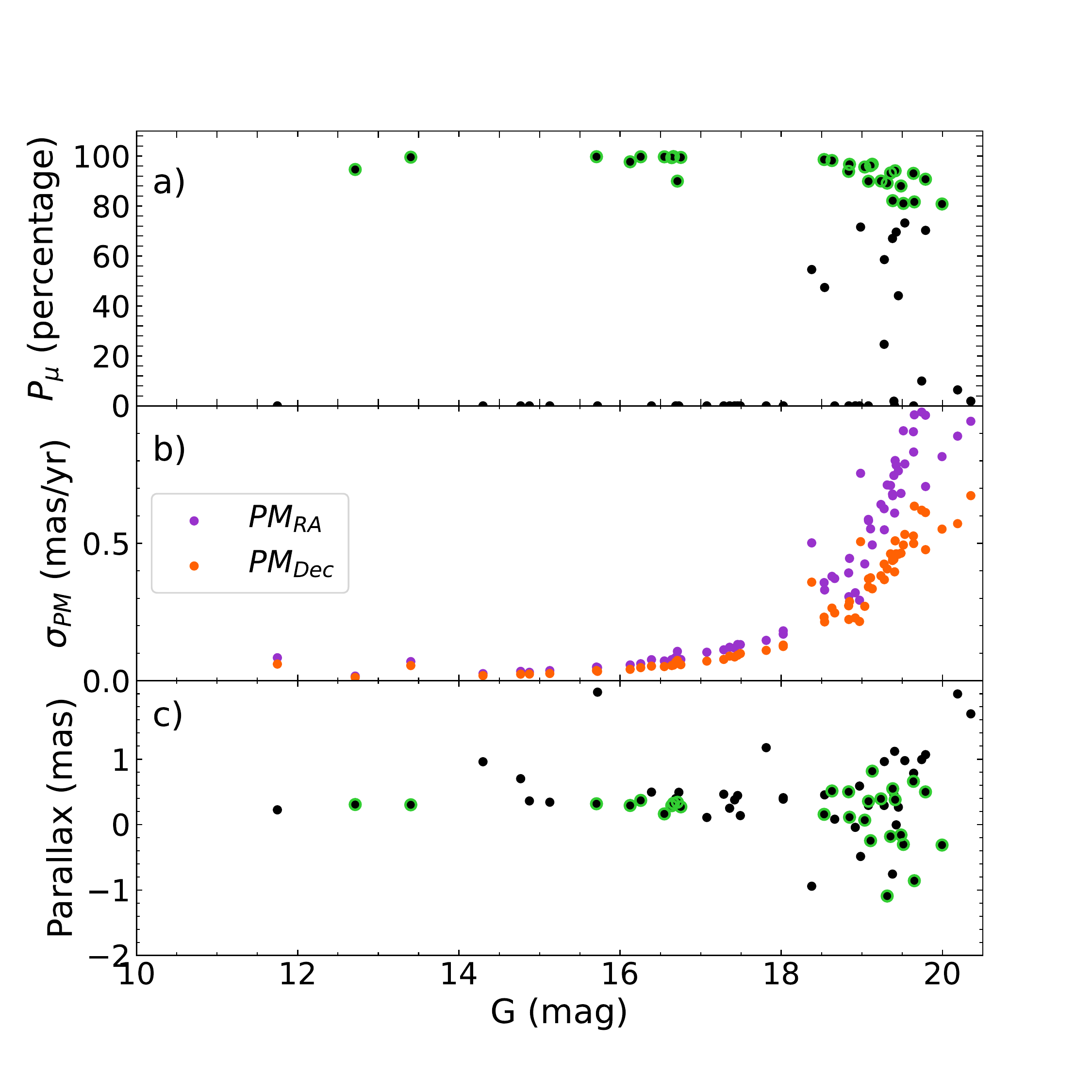}
\end{minipage}
\caption{PM VPDs (upper left panel) and \emph{Gaia DR3 G} vs. $(G_{BP}-G_{RP})$ CMDs for the stars located within the hull of the E70 bubble. The subpanel 'a)' in both the left panels, is for all the stars; whereas the subpanels 'b)' represent probable cluster and field stars, respectively. The Right panel is representing membership probability $P_{\mu}$, PM errors $\sigma_{PM}$, and parallax of the stars as a function of \emph{Gaia DR3 G} located within the hull of E70 bubble. A total of 29 stars having $P_{\mu} \geq 80\%$, are considered as cluster stars and shown by green ringed black circles.}
     \label{fig:gaiadr3}
\end{figure*}

Proper motion (PM) is among the best parameters to understand the structure and membership probability of a cluster. \emph{Gaia DR3} data which provides precise parallax up to faint limits ($G\approx21$), has been used for the same in E70 bubble. Gaia data located within the hull and having PM error $\leq$ 1 mas\,yr$^{-1}$, has been used for the determination of membership probability. The upper left panel of Figure \ref{fig:gaiadr3} represents the vector-point diagram (VPD) of PM errors; $\mu_{\alpha}cos\delta$ and $\mu_{\delta}$ whereas the lower left panel represents the analogous G versus $G_{BP} - G_{RP}$ CMDs. We have found a prominent clump centred at ($-$0.33, $-$1.77) mas\,yr$^{-1}$ and is of the radius 0.5 mas\,yr$^{-1}$. The stars outside this circle, are considered as field stars. The probable cluster stars are following well defined main sequence (MS) in the CMD whereas the probable field stars are showing a broad distribution in CMD. We have considered that the cluster is situated at a distance of 3 kpc and radial velocity of the dispersion is 1 \kms, which gives the dispersion in PM ($\sigma_c$) $\approx$ 0.07 mas\,yr$^{-1}$. The field stars are centred at (0.97 $\pm$ 3.15, $-$3.01 $\pm$ 5.03) mas\,yr$^{-1}$. By using these values, we have calculated the frequency distribution of cluster stars ($\phi^{\nu}_c$) and field stars ($\phi^{\nu}_f$), as determined in \citet{2020MNRAS.498.2309S}.

The membership probability of the i$^{th}$ star is calculate by:

\begin{equation}
    \begin{split}
        P_{\mu}(i) = \frac{n_c \times \phi^{\nu}_c(i)}{n_c \times \phi^{\nu}_c(i) + n_f \times \phi^{\nu}_f(i)}
    \end{split}
\end{equation}

Here $n_c$ (= 0.28) and $n_f$ (= 0.72) represent the normalized number of cluster and field stars, respectively. The stars having membership probability ($\geq$ 80) are considered as the probable members of the cluster which extend up to 20 mag in G band. The above made us able to calculate the membership probability of all the stars in the E70 bubble. The right panel of Figure \ref{fig:gaiadr3} represents the PM errors $\sigma_{PM}$ and parallax of the stars as a function of \emph{Gaia DR3 G} located within the hull of E70 bubble. A total of 29 stars having $P_{\mu} \geq 80\%$, are considered as cluster stars.

\bibliography{bibliography}{}
\bibliographystyle{aasjournal}

\end{document}